\documentclass[preprint]{aastex}
\usepackage{emulateapj5}
\def\stacksymbols #1#2#3#4{\def\theguybelow{#2}
	\def\verticalposition{\lower#3pt}
	\def\spacingwithinsymbol{\baselineskip0pt\lineskip#4pt}
	\mathrel{\mathpalette\intermediary#1}}
\def\intermediary #1#2{\verticalposition\vbox{\spacingwithinsymbol
	\everycr={}\tabskip0pt
	\halign{$\mathsurround0pt#1\hfil##\hfil$\crcr#2\crcr
		\theguybelow\crcr}}}
\def\lta{\stacksymbols{<}{\sim}{2.5}{.2}}
\def\gta{\stacksymbols{>}{\sim}{3}{.5}}

\begin{document}

\title{HEATED COOLING FLOWS}

\author{Fabrizio Brighenti$^{1,2}$ \& William G. Mathews$^1$}

\affil{$^1$University of California Observatories/Lick Observatory,
Board of Studies in Astronomy and Astrophysics,
University of California, Santa Cruz, CA 95064\\
mathews@lick.ucsc.edu}

\affil{$^2$Dipartimento di Astronomia,
Universit\`a di Bologna,
via Ranzani 1,
Bologna 40127, Italy\\
brighenti@bo.astro.it}






\vskip .2in

\begin{abstract}

In conventional models of galactic and cluster cooling flows  
widespread cooling (mass dropout) is assumed to 
avoid accumulation of unacceptably large central masses. 
This assumption is often criticized because 
the observed mass of cooled gas and newly formed stars
are much less than the cooling rates imply. 
Moreover, recent XMM observations have failed 
to find spectral evidence for locally {\it cooling} gas. 
This remarkable discovery 
has revived the notion that cooling flows do  
not cool in the manner previously assumed, 
but are instead heated by some process such as an 
intermittent or low-level AGN in the central galaxy. 
To explore this hypothesis,
we consider the gas-dynamical consequences of heated
galactic cooling flows under many different 
heating scenarios without specifying the detailed physics of the 
heating process. 
We have been unable to find a single acceptable 
heated flow in reasonable agreement 
with well-observed hot gas temperature and density profiles.
Even for poor fits to the observations,
these models require finely tuned heating scenarios. 
Idealized flows in which radiative cooling is perfectly 
balanced by global 
heating are grossly incompatible with observations.
Flows heated by central
feedback often result in eposidic cooling 
that is associated with quasi-cyclic changes in the 
hot gas density profile;
at present there is no
observational evidence for this. 
Paradoxically, when cooling flows 
are centrally heated (or pressurized), 
they experience spontaneous non-linear compressions that 
result in spatially widespread cooling instabilities. 
Therefore, spectral evidence for cooling gas is expected
in either traditional cooling flows with {\it ad hoc} mass dropout
or in heated cooling flows without the dropout assumption.

\end{abstract}



\section{INTRODUCTION}

Early results from the XMM and Chandra X-ray telescopes 
have confounded and intensified 
long-simmering controversies regarding cooling flows
in elliptical galaxies and galaxy clusters.
In the absence of detailed spectroscopic data, 
it had been assumed that the virialized hot gas 
in these systems cools over a large volume as the gas 
slowly flows 
inward, depositing several to many 100 $M_{\odot}$ yr$^{-1}$.
Since optical evidence for normal star formation 
is limited, it is usually assumed that the cooled gas 
forms preferentially into stars of low mass.
Indeed, there is some evidence for cooled gas 
in H$\alpha$ + [NII] emission in cooling flows 
(e.g. Heckman et al. 1989; 
Caon, Macchetto \& Pastoriza 2000) and 
for populations of younger stars 
(McNamara 1997; Cardiel et al. 1998).
Gas dynamical models with spatially distributed 
mass dropout (cooling) have been devised 
that agree with most observations.
Many X-ray observers have claimed that 
cooling flows can be understood if mass drops
out so that the remaining mass flow varies as 
${\dot M}(r) \propto r$ 
(e.g. Thomas, Fabian, \& Nulsen 1987). 

The unexpected result from the Reflection Grating Spectrometer 
(RGS) on XMM is that the multiphase spectral 
signature of gas cooling to low temperatures has not been seen in 
Abell 1835 (Peterson et al. 2001) or 
Abell 1795 (Tamura et al. 2000). 
XMM/EPIC spectra of M87/Virgo also fail to reveal
evidence of cooling at the expected rate
(B\"ohringer et al. 2001a, b, c; Molendi \& 
Pizzolato 2001; Matsushita et al. 2002), 
in contradiction to the long-standing 
observation with {\it Einstein} FPCS by Canizares et al. (1982),
recently confirmed by Buote (2001).
High resolution XMM/RGS spectra 
of elliptical galaxy NGC 4636 show strong emission 
from the OVIII Ly$\alpha$ line at 0.654 keV 
characteristic of hot gas at $T \gta 6 \times 10^6$ K, but the 
OVII resonance line at 0.574 keV, expected from 
cooling gas at $T \sim 2 \times 10^6$ K,  
is strangely absent (Xu et al. 2001).
On the other hand,
FUSE observations of NGC 4636 detect OVI
emission from gas at $T \sim 3 \times 10^5$ K indicating a 
mass cooling rate ${\dot M}$ near (but less than) 
that predicted by conventional multiphase
cooling flow models (e.g. Bregman et al. 2001).
Nevertheless, as a result of the XMM data,  
an increasing number of theorists and observers
are doubting the reality of traditional cooling flows 
and radically modified alternative 
theoretical models have been proposed 
(Fabian et al. 2001; Morris \& Fabian 2001).

One of the most consistent and (initially) appealing 
suggestions is that the gas in 
these so-called cooling flows is not cooling by 
radiation losses because it is 
continuously or intermittently heated by energy released 
near supermassive black holes in the cores of central 
E or cD galaxies. 
This seems reasonable since 
almost every giant E or cD galaxy is known to be a source 
of AGN activity at some level, ranging from the 
barely observable to the most powerful known radio galaxies. 
It has been suggested that 
the hot thermal plasma could be heated by AGN energy 
in the form of shock waves,
Compton scattering or by dissipation of the 
energy in relativistic particles. 
Certainly some energy transfer to the gas must be occurring, 
and the AGN is likely to operate in a feedback mode, 
supplying energy whenever gas cools at the center of 
the cooling flow and fuels the black hole. 
Enough AGN energy is available to significantly 
influence the flow, provided it can be 
judiciously transfered to the gas.

Although the heating hypothesis is appealing,
we show in the discussion below 
that unforeseen complications 
rule out many -- and perhaps all -- heating scenarios 
that affect the gas flow significantly.
Our objective here is to study the time dependent 
evolution of cooling flows that are heated in a variety 
of different ways.
We do not consider the detailed and uncertain 
microphysics of the heating process.
Instead, we assume that heating occurs according to 
various space and time dependent recipes as described 
by a few adjustable parameters. 
For success, a heated flow must agree reasonably well with 
the density and temperature distributions observed in 
cooling flows.
In addition, this agreement must be achieved with a 
wide range of heating parameters so that it is not necessary 
to ``fine-tune'' the heating process, which may be 
inherently stochastic.
Furthermore, the mass of gas accumulated at 
or near the center of the flow must 
be consistent with the masses of central black holes and 
stellar mass to light determinations from central velocity 
dispersions. 
Finally, 
to be fully compatible with the lack of evidence for
intrinsic multiphase gas in XMM X-ray spectra, 
heated cooling flows should have 
little or no distributed cooling.

We focus here on galactic or galaxy group scale 
cooling flows associated with a single central 
massive elliptical galaxy. 
Indeed, all giant E galaxies are thought to have formed
by mergers of smaller galaxies in group environments.
As a result, most luminous E galaxies are surrounded
by extended X-ray halos confined by the group potential, 
containing more hot gas than can be accounted for by
stellar mass loss in the central galaxy.
Within the central galaxy, however, new 
gas is continuously supplied by stellar mass loss 
so the absence of observed 
cooling flow dropout is consequently more of a problem.
Since larger cluster cooling flows 
invariably flow into a central E or cD 
galaxy, our arguments generally apply for these larger flows 
as well, although more central heating would be required 
to decelerate the inflow velocity in cooling in cluster flows.

We deliberately choose the simplest possible heating scenarios;
if these do not give satisfactory results, it is unlikely 
that adding details will help.
We warn readers that most or all of
the heated cooling flow
models described here are found to be unacceptable by the 
criteria that we have set.

It is particularly interesting that 
multiphase cooling over a large 
region of the flow is a frequent and 
natural consequence of heating or 
additional nonthermal pressure near the center of the galaxy.
Centrally heated or pressurized gas 
decelerates (or reverses) the normal cooling inflow in the 
galactic core. 
At the radius where normally
inward flowing gas from large radii encounters stagnant
or outflowing gas in the core, a nonlinear
compression develops, resulting in thermal instabilities.
Since this cooling instability is 
inherently nonlinear, it was not 
found in linear studies (e.g. Balbus \& Soker 1989).
However, the expected multiphase spectral signature for 
such cooling gas is not yet reported by XMM observers.

AGN heating in cooling flows is not a new idea. 
Before we describe our own models, we review 
in the following section some of the 
heated models that have already been proposed.

\section{A Brief History of Heated Cooling Flows}

\subsection{Possible Heating Mechanisms}

\subsubsection{Thermal Conduction}

The positive temperature gradient ($dT/dr > 0$) in 
the central regions of cooling flows is 
due to radiative cooling and, on galactic scales, 
to mixing with cooler stellar mass loss. 
Since the heat flux opposes the thermal gradient, 
it has been suggested that heat could be conducted inward 
toward the center of the flow, balancing the radiative cooling 
there and arresting or retarding the 
global evolution of the cooling flow and its mass dropout
(Stewart, Canizares, Fabian \& Nulsen 1984;
Bertschinger \& Meiksin 1986; Meiksin 1988; 
Bregman \& David 1988; 
Loewenstein, Zweibel \& Begelman 1991;
Brighenti \& Mathews 2002).
Unfortunately, these attempts 
to avoid radiative cooling have been generally unsuccessful 
because conductive heat flux and radiation losses are important 
in different flow regions.
The Spitzer conductivity depends strongly on temperature 
($\kappa \propto T^{5/2}$) and is often very small at the 
centers of cooling flows where radiative losses ($\propto n^2$) 
are greatest. 
Furthermore, when the temperature is low  
everywhere in the flow, $T \lta 1$ keV, as in 
galactic scale cooling flows, 
thermal conductivity is small at every radius.
Thermal conduction can be reduced by tangled magnetic 
fields, which may explain the existence of ``cold fronts'' 
in some cluster cooling flows 
(Ettori \& Fabian 2000, but also see 
Rosner \& Tucker 1989; Narayan \& Medvedev 2002).
For galaxy clusters, the inward conduction of 
thermal energy can slow or stop the 
cooling flow, but the conductivity must be carefully adjusted 
to fit a particular observed flow.
Generally the results are 
very sensitive to the conduction reduction factor 
and are cluster-dependent, 
a form of fine-tuning that invalidates most models.

\subsubsection{Heating by Cosmic Rays}

Radiofrequency emission from cosmic ray electrons is observed 
in the galactic centers of most or all cooling flows.
If the less visible proton cosmic ray 
component is scaled as in the 
Galaxy, Coloumb collisions could produce 
substantial heating (Rephaeli 1987; Rephaeli \& Silk 1995). 
Hints that this type of heating could be important are 
just now emerging. 
By combining the most recent X-ray and radio data 
from the X-ray holes in NGC 1275 (Perseus cluster), 
Fabian et al. (2002) find evidence for an energy 
density of suprathermal particles that exceeds that of the 
fast electrons by $(180 - 500)f$, where $f$ is the (unknown) 
volume filling factor of the electron component. 
In addition, the flattened gas cores at $r \lta 700$ pc 
in bright Virgo E galaxies strongly suggests the 
presence of an additional non-thermal 
pressure component which could 
be due to (non-radiating) cosmic rays or magnetic fields 
(Brighenti \& Mathews 1997a). 
More commonly, cosmic ray heating is assumed to be due 
dissipation of Alfven waves excited by relativistic electrons 
or protons streaming along field lines. 
This possibility has been widely discussed in previous 
work with somewhat inconclusive results 
(Lerche 1967; Kulsrud \& Pearce 1969; Wentzel 1969; 
Skilling 1971; Scott et al. 1980; Tucker \& Rosner 1983;
Rose et al. 1984; 
Rosner \& Tucker 1989; Begelman \& Zweibel 1994). 

\subsubsection{Inverse Compton Heating by hard AGN Radiation}

Compton heating by isotropic quasar radiation 
(Binney \& Tabor 1995; Ciotti \& Ostriker 2001) 
is effective if the Compton temperature  
$T_C \approx \langle h \nu \rangle /4$ exceeds the virial
temperature, typically $T_{vir}\sim 1$ keV ($\sim 10^7$ K) 
for galactic cooling flows. 
Gamma ray observations with EGRET since 1991 have 
discovered $\sim 50$ flat spectrum blazars 
(BL Lac AGN and OVV type QSOs) 
in which $T_C \sim 10^8 - 10^9$ K due to the dominance of 
gamma rays in the energy output 
(von Montigny et al. 1995;.
Impey 1996; Hartman et al. 1999).
If all normal elliptical galaxies 
emitted such radiation isotropically following 
short-lived accretion events, 
cooling flow gas near the AGN, when 
heated to $T_C \gg T_{vir}$, would 
violently expand into the outer flow, producing strong 
shock waves. 
Outward propagating shocks could heat the entire cooling 
flow provided the gas density has a steep radial 
dependence, $n \propto r^{-s}$ with $s \ge 2$ as in 
the models of Ciotti \& Ostriker (2001).

However, the (generally high redshift) EGRET sources exhibit 
large amplitude, random time variability.
This variability is thought to occur because 
the radiation is beamed and Dopper boosted in a jet 
directed along our line of sight. 
This interpretation is supported by the association of 
gamma-dominant EGRET blazars with radio loud, flat spectrum 
central VLBI sources having large superluminal jet velocities.
None of the EGRET sources are normal Seyfert galaxies,
radio quiet QSOs or E galaxies 
except possibly Cen A (Hartman et al. 1999).
Therefore, the assumptions that EGRET sources 
are isotropic gamma ray emitters and exist in all 
luminous elliptical galaxies are in doubt.
A highly collimated beam of 
gamma radiation would heat the cooling flow 
much less efficiently.
Nulsen \& Fabian (2000) claim that the Compton temperature 
of more typical isotropic AGNs and QSOs is less than $10^7$ K,
so cooling flow gas is more likely to be Compton cooled 
than heated.
Nevertheless, it may not be possible to rule out 
isotropic bursts of gamma rays provided the duty cycle 
is sufficiently short.
If so, the Compton heating would be quite localized, 
decreasing as $r^{-2}$ from the AGN.

\subsubsection{Shock Waves from Expanding Radio Plasma}

If the energy density in radio lobes is sufficiently 
large, they will expand supersonically,  
producing shocks that heat the cooling flow gas
(Clarke, Harris, \& Carilli 1997; 
Heinz, Reynolds, \& Begelman 1998; 
Begelman \& Cioffi 1989;
Rizza et al. 2000;
Reynolds, Heinz, \& Begelman 2001;
Soker, White, David \& McNamara 2001). 
While such heating probably occurs at some level, 
recent Chandra and XMM observations show little evidence 
that the radio sources are actually heating the bulk 
of the gas or even gas just adjacent to the 
radio lobes (Fabian 2001; McNamara 2001).

\subsection{Recent Cooling Flow Models with Heating}

Loewenstein, Zweibel \& Begelman (1991) 
discuss steady state 
cooling flows that are heated by cosmic rays in 
the center and by thermal conduction from larger radii.
While their model is successful in producing 
central regions with $dT/dr > 0$, 
the central gas density gradient $dn/dr$ 
is flatter than observed 
because the gas is both supported and heated by cosmic rays. 
They conclude that cosmic-ray heating cannot balance 
radiative cooling in intracluster cooling flows.

Citing the difficulties encountered in producing thermal 
instabilities in nearly hydrostatic cooling flows 
(Malagoli, Rosner \& Bodo 1987; Balbus \& Soker 1989;
Hattori \& Habe 1990; Malagoli, Rosner \& Fryxell 1990), 
Tabor \& Binney (1993) assume 
that gas cannot cool in a spatially distributed fashion,
and devise steady state flows 
in which a convective core heated from the center 
is attached to an outer cooling inflow.
Unfortunately, in the convective core the temperature gradient 
is determined by the gravitational potential, 
$(k /\mu m_p)(dT/dr) = -(2/5)(d \Phi /dr)$, 
and must always be negative, contrary to observation.
Furthermore, the net gas velocity within the convective 
core is everywhere negative, causing all the gas 
shed from evolving stars to cool at the very center. 
After a Hubble time 
the mass of centrally cooled gas $\sim 10^{10}$ $M_{\odot}$
is unacceptably large.
Binney \& Tabor (1995) discuss time dependent, 
centrally heated models in which distributed mass 
dropout is not permitted, except perhaps near the core.
These models are heated by a large number of Type Ia 
supernovae at 
early times which can drive a galactic outflow 
throughout the evolution  
but which also enrich the hot gas with iron 
far in excess of observed abundances. 
Strong central supernova heating results in 
$dT/dr < 0$ at every time in the flow evolution, 
unlike observed profiles. 
Their best-fitting models for NGC 4472, 4636 and 4649 
agree with the 
observed gas density profiles only during a brief 
$\sim 10^8$ year period just before a cooling catastrophe 
when the central gas density increases rapidly due to 
radiative losses. 
In additional models, Binney \& Tabor include 
intermittent AGN heating within $\sim 1$ kpc triggered 
by gas inflow into the center.
As SNIa heating subsides and the flow evolves toward a 
central cooling catastrophe, the AGN heating is activated 
but the central gas density still continues to increase, 
and, as Binney \& Tabor suggest, 
this may result in distributed cooling and 
star formation in the central 1-2 kpc. 
By the standards we set for success in our heated models, 
this flow would be regarded as a failure because cooling 
within a few kpcs would be visible as a multiphase 
component in XMM spectra.

Tucker \& David (1997) study time dependent models 
of cluster scale heated flows without cooling dropout or 
mass supply from stars.
Gas near a central AGN is intermittently heated 
by relativistic electrons when gas flows into the origin.
For these models, the temperature and entropy gradients 
are both negative while the density gradient is positive;
such flows are both convectively and Rayleigh-Taylor unstable.
When Tucker \& David include thermal conduction, 
the outward heat flux results in stable 
solutions having cores of nearly flat density,
but which are nearly isothermal, in disagreement with 
observations. 
In solutions with thermal conduction  
the mass flow ${\dot M}$ into 
the central regions is lower, and 
${\dot M}(r)$ slowly varies 
for many Gyrs as a long-lived transient.

If gas in isolated elliptical galaxies is locally 
Compton heated by powerful isotropic gamma rays 
($L_{bol} \sim 10^{46} - 10^{48}$ erg s$^{-1}$; 
$T_C \sim 10^9$ K) 
following mass accretion at the center, 
strong outward moving shocks can heat gas out to 
$\sim 5$ kpc and beyond (Ciotti \& Ostriker 2001). 
The AGN-QSO continuum is only activated about 
$\sim 10^{-3}$ of the time, so only one in $\sim 1000$ of 
known elliptical galaxies should be a gamma-dominant quasar 
at any time.
For these models Ciotti \& Ostriker assume
a high Type Ia supernova rate and no supply of gas
by secondary infall. 
Such cooling flows, which are less bound 
and have small inertial masses, 
can be significantly modified by
the additional AGN heating.
In these models gas is driven out of the galactic core by 
shocks, 
so distributed cooling dropout is not necessary nor 
do large masses of gas cool at the galactic core; 
these are desirable attributes. 
As with previous calculations with heating,
however, $dT/dr$ is generally negative and  
the gas density is has a broad central core. 
The density and temperature profiles 
vary depending on the recent history of 
heating events during the previous sound-crossing time.
In their preferred model 1 the 
central core in the X-ray surface 
brightness distribution $\Sigma_x$ varies 
in size from 1 to 40 kpc which is much larger than 
the $\sim 0.6$ kpc core in NGC 4649, 
an elliptical with an $L_B$ similar to that 
used in the Compton heated models. 
Because of radially propagating shock waves, 
the X-ray images of these models are considerably 
more irregular than 
$\Sigma_x$ profiles observed in elliptical galaxies.
As in the models by Binney \& Tabor (1995), 
the X-ray luminosity $L_x$ is expected to be large
during those phases of the heating cycle when the X-ray 
size $r_x$ is small and the X-ray image is more concentrated.
But observations show just the opposite: 
$r_x$ and $L_x$ are positively correlated 
(Mathews \& Brighenti 1998). 
Finally, such 
large excursions in the $L_x$-$r_x$ plane following 
explosive heating events would be greatly reduced in 
more realistic cooling flows 
having larger masses of gas 
acquired from secondary infall, 
which extends far beyond the optical galaxy. 

\section{Criteria for a Successful Heated Flow}

We seek computed solutions for heated cooling flows that satisfy 
three criteria.

\noindent
(1) After evolving for $\sim 13$ Gyrs, the 
hot gas density and temperature profiles should resemble 
those observed in massive E galaxies. 
For this purpose 
we adopt NGC 4472 as a representative luminous E galaxy 
for which the density and temperature are well-determined from 
X-ray observations.

\noindent
(2) When our models are computed to $t = 13$ Gyrs, 
assumed to be the current time, 
the rate of cooling (mass dropout) 
should not exceed cooling rates ${\dot M}$ 
observed in typical bright E galaxies. 
Unfortunately at the present time only a few 
galactic scale cooling flows have been observed 
with XMM and the indicated values of ${\dot M}$ 
based on spectral features differ quantitatively. 
Nevertheless, 
X-ray observers all agree that ${\dot M}$ is much lower than 
predictions of standard cooling flow models.
From their XMM-EPIC observations of 
M87 Molendi \& Pizzolato (2001) 
find ${\dot M} < 0.5$ $M_{\odot}$ 
yr$^{-1}$ while $40 \pm 5$ $M_{\odot}$
yr$^{-1}$ and $\sim 10$ $M_{\odot}$ yr$^{-1}$ 
were predicted by Peres et al. (1998) and 
Fabian et al. (1984) respectively.
More recently Matsushita et al. (2002) claim to have 
detected a total cooling rate in M87 within 20 kpc 
of $0.8$ $M_{\odot}$ yr$^{-1}$ while 
8 $M_{\odot}$ yr$^{-1}$ was expected in their 
traditional cooling flow model 
inside the same radius, but the statistical likelihood 
of this solution is small. 
Recent XMM observations of the E galaxy 
NGC 5044 fail to find spectroscopic 
evidence of cooling below $\sim 1$ keV, implying an 
${\dot M}$ consistent with zero 
(Buote et al. 2002).
For NGC 4636, which more closely resembles our fiducial 
galaxy NGC 4472, the XMM-RGS observations of Xu et al. (2001)
can only provide upper limits to the total mass cooling 
rate. 
Xu et al. find a complete absence of emission lines from gas 
cooling below 0.5 keV. 
For an upper limit they find 
${\dot M} < 0.2 - 0.3$ $M_{\odot}$ yr$^{-1}$, 
which is 3-5 times less than cooling rates based on 
X-ray imaging, $\sim 1$ $M_{\odot}$ yr$^{-1}$.
However, FUSE observations of the OVI doublet by 
Bregman, Miller \& Irwin (2001) indicate that gas in NGC 4636 is 
cooling through temperatures $T \sim 3 \times 10^5$ K 
at a rate ${\dot M} = 0.43 \pm 0.06$ $M_{\odot}$ yr$^{-1}$. 
In view of these inconsistencies in the current observations, 
we determine the success of a computed heated cooling flow in 
reducing ${\dot M}$ by comparing with the cooling rate 
of an identical galaxy flow model without heating. 
Guided by the X-ray observations, the ${\dot M}$ of a heated 
flow should be reduced by about an order of magnitude,
${\dot M} \lta 0.1$ $M_{\odot}$ yr$^{-1}$, to 
be compatible with XMM observations.

\noindent
(3) Finally, when our heated cooling flow models are computed 
to the current cosmic time, $t \sim 13$ Gyrs, the total mass 
of cooled gas within any radius must not exceed the combined mass 
of the stars and central black hole 
as determined from the stellar velocity dispersion.
The black hole mass 
is $\sim 2.6 \times 10^9$ $M_{\odot}$ for NGC 4472 
(Magorrian et al. 1998). 
The stellar mass within about 200 pc is equal to the 
black hole mass.
At a radius of 100 pc, within which much of the gas cools 
in many of our galactic flow models, 
the combined stellar and black hole 
mass in NGC 4472 (as described below) 
is $M(r<100~{\rm pc}) \sim 2.9 \times 10^9$ $M_{\odot}$. 

\section{EQUATIONS}

Our reference galaxy NGC 4472 has a 
stellar mass $M_{*t} = 7.26 \times 10^{11}$
$M_{\odot}$, assuming $M/L_B = 9.2$ and a distance of 
17 Mpc.
If the X-ray luminosity of NGC 4472 
$L_x(0.5 - 2~{\rm keV}) = 4.54 \times 10^{41}$ erg s$^{-1}$ 
is assumed to have 
been constant over $t_n = 13$ Gyrs and there 
had been no heating, 
the total mass of cooled interstellar gas would be 
$M_{cool} \approx
t_n L_x/(5 k T / 2\mu m_p) \approx
4 \times 10^{10}$ $M_{\odot}$. 
If there is no heating, the current rate of cooling 
in NGC 4472 is 
$\sim 1.2$ $M_{\odot}$ yr$^{-1}$ (see below).

For galactic scale cooling flows, the usual gas dynamical 
equations must include appropriate source and sink terms: 
\begin{equation}
{ \partial \rho \over \partial t}
+ {\bf \nabla} \cdot \rho {\bf u} 
= \alpha \rho_*
- q{\rho \over t_{cool}},
\end{equation}
\begin{equation}
\rho \left[ { \partial {\bf u} \over \partial t}
+ ({\bf u} \cdot {\bf \nabla}){\bf u} \right]
= - {\bf \nabla} P
- \rho {\bf \nabla}\Phi - \alpha \rho_* {\bf u},
\end{equation}
and
$$\rho \left[ { \partial \varepsilon \over \partial t}
+ ({\bf u} \cdot {\bf \nabla})\varepsilon \right]
= {P \over \rho}
\left[ { \partial \rho \over \partial t}
+ ({\bf u} \cdot {\bf \nabla})\rho \right]
- { \rho^2 \Lambda \over m_p^2}$$
\begin{equation}
+ \alpha \rho_*
\left[ \varepsilon_o - \varepsilon - {P \over \rho}
+ {u^2 \over 2} \right] 
- {\bf \nabla} \cdot {\bf F}_{conv} 
- {\bf \nabla} \cdot {\bf F}_{cond}
+ H(r,t) \rho
\end{equation}
where $\varepsilon = 3 k T / 2 \mu m_p$ is the
specific thermal energy, 
$\mu = 0.62$ is the molecular weight and $m_p$ the proton mass.
The terms involving $\alpha = \alpha_* + \alpha_{sn}$
represent the addition of 
mass, momentum and energy to the hot gas 
by stellar mass loss and Type Ia supernova;
usually $\alpha_{sn} \ll \alpha_*$.
The stellar mass loss varies as
$\alpha_* = \alpha_*(t_n)(t/t_n)^{-1.3}$ s$^{-1}$ 
where $\alpha_*(t_n) = 4.7 \times 10^{-20}$ s$^{-1}$ 
(e.g. Mathews 1989).

The gravitational potential $\Phi$ in the momentum equation 
includes contributions from both stellar and dark mass.
The de Vaucoulers stellar density $\rho_*(r)$ and mass 
$M_*(r)$ profiles are uniquely determined 
by the total mass
$M_{*t} = 7.26 \times 10^{11}$
$M_{\odot}$ and effective radius $r_e = 8.57$ kpc
(e.g. Mellier \& Mathez 1987).
We assume a flat stellar core inside the break radius 
$r_b = 0.205$ kpc (Faber et al. 1997), 
which contains less than one percent of the total 
stellar mass. 
The NFW halo in NGC 4472 has a current virial mass 
$M_{h,vir} = 4 \times 10^{13}$ $M_{\odot}$
based on its extended X-ray image 
(Brighenti \& Mathews 1997). 
For the simple $\Omega = 1$ cosmology we consider 
with Hubble constant $H_o = 50$ km s$^{-1}$ Mpc$^{-1}$,
the current dark halo density distribution is 
$$ {\rho_h \over \rho_c}  = 
{200 \over 3}{ c^3 \over
\ln (1 + c) - [c/(1+c)]}
\cdot {1  \over 
cx(1 + cx)^2}$$
where $\rho_c = 3 H_o^2 /8 \pi G$ is the 
critical density,
$c = 498 [M_{h,vir}/M_{\odot}]^{-0.1233}$ 
is the concentration, $x = r/r_{h,vir}$, 
and $r_{h,vir} = (3 M_{h,vir} / 4 \pi 200 \rho_c)^{1/3}$
(Navarro, Frenk \& White 1996).
While the large halo mass $M_{h,vir}$ is appropriate for 
the galaxy group from which NGC 4472 formed, 
our results are surprisingly insensitive to the (uncertain) 
value of $M_{h,vir}$.
For example, when the same flow is computed with a 
halo mass reduced by ten, 
$M_{h,vir} = 4 \times 10^{12}$ $M_{\odot}$, 
we typically find that the 
gas temperature and density for $r \sim 10 - 50$ kpc are 
reduced by only about 15 and 45 percent respectively. 
Since our assessment of the success of heated cooling flow  
models is weighted toward agreement with observations of NGC 4472 
within the bright optical galaxy, $r \lta 20$ kpc, 
these relatively small changes in the flow models 
due to the uncertain value 
of $M_{h,vir}$ have no effect on our conclusions. 

Guided by the relative sizes of the optical and 
X-ray images for E galaxies (Mathews \& Brighenti 1998), 
we consider two types of evolutionary calculations 
for the hot galactic gas: 
``isolated'' and cosmologically ``accreting''. 
Accreting models have large halos of X-ray gas acquired from 
cosmic flow into the galaxy group potential 
in which the central E galaxy formed. 
Alternatively, 
in our isolated E galaxy models all of the interstellar 
gas is supplied only by stellar mass loss and supernovae. 
In isolated galaxies the X-ray and optical images are of comparable 
size. 
In spite of our designation ``isolated,'' 
(typically non-central) E galaxies with comparable 
X-ray and optical sizes 
probably lack hot circumgalactic gas either because 
of tidal stripping or because they never received their 
full allocation of cosmic gas.
For simplicity, we assume 
that isolated E galaxies evolve from 
$t = t_i = 1$ Gyr 
to the current time $t_n = 13$ Gyrs with 
stellar and dark matter potentials fixed at their 
current values. 
In these models we begin the calculation by 
filling the galaxy with a very small gas density 
($n = 10^{-6}$ cm$^{-3}$) at or 
near the virial temperature ($T = 10^7$ K). 
Shortly after the calculation begins the mass that flows 
from the stars overwhelms the initial gas 
and the final solution is independent 
of the initial gas. 
Accreting models are assumed to evolve from $t_i$ in a 
simple flat universe, $\Omega = 1$ 
with Hubble constant $h = H/100 = 0.5$
and $\Omega_b = 0.05$, in which the 
dark halo grows from the inside out, beginning with 
a tophat cosmic perturbation 
(see Brighenti \& Mathews 1999 for further details). 
In these models the continuing infall of cosmic gas 
creates hot gas distributions that extend far beyond the 
optical image.
These flows approximate E galaxies at the center of groups.

The injection energy in Equation (3) is 
$\varepsilon_o = 3 k T_o / 2 \mu m_p$ where
$T_o = (\alpha_* T_* + \alpha_{sn} T_{sn})/\alpha_*$
is the mean temperature of gas expelled by stellar mass 
loss and supernovae.
The supernova term 
$$\alpha_{sn} T_{sn} = 2.13 \times 10^{-8} {\rm SNu}(t) 
\left( {E_{sn} \over 10^{51}~{\rm ergs}}\right) 
h^{-1.7} L_B^{-0.35}
~~~{\rm K}~{\rm s}^{-1}$$
is proportional to the Type Ia rate in SNu units, 
SNu$(t) = 0.06(t/t_n)^{-1}$ supernovae in 100 years 
per $10^{10} L_B$. 
This value is consistent with the poorly known 
observed rate, $\sim 0.08 \pm 0.03$ SNu 
from Cappellaro \& Turato (2000).
Since $\alpha_{sn} T_{sn}$ is generally much larger
than $\alpha_* T_*$, we simply assume 
$T_* = (\mu m_p / k) \sigma_*^2$ with a constant 
stellar velocity dispersion 
$\sigma_* = 230$ km s$^{-1}$; 
the $\alpha_* T_*$ term has essentially no effect 
on our computed flows. 
We evaluate $\alpha_{sn} T_{sn}$ 
using $E_{sn} = 10^{51}$ ergs; further details concerning 
the expression above for $\alpha_{sn} T_{sn}$ can 
be found in Brighenti \& Mathews (1999). 
Unless otherwise noted,
$\alpha_*(t)$, $\alpha_{sn}(t)$ and $M_{*t}$ are the 
same for all models and 
conduction and convection 
are zero, ${\bf F}_{cond} = {\bf F}_{conv} = 0$.
If desired, we can include an additional 
heating source $H(r,t)$ in Equation (3) as described in 
more detail below.

The dimensionless coefficient $q$ in the sink term 
in Equation (1) can be 
used to remove hot gas from the flow over some large volume
or to remove gas that has already cooled. 
In either case 
$-q\rho /t_{cool} \propto -q\rho^2,$
where $t_{cool} = 5 m_p k T / 2 \mu \rho \Lambda$
is the time for gas to cool locally by radiative
losses at constant pressure. 
For simplicity we evaluate the cooling coefficient
$\Lambda(T)$ using the results
of Sutherland \& Dopita (1993) assuming solar
abundance.
The cooling rate and the gas density within
$\sim r_e$ are slightly underestimated
in central galaxies where $z_{Fe} \gta z_{Fe,\odot}$.

Ideally, if the gas in heated cooling flows does not cool, 
there would be no reason for this {\it ad hoc} sink term.
However, in the heated cooling flows 
models discussed below we find that gas cools to low 
temperatures in localized regions 
(drips) where radial compression 
(${\bf \nabla} \cdot {\bf u} < 0$) is sustained.
To remove the accumulated cooled gas from the flow 
in a numerically satisfactory manner in 
spherical 1D flows we let 
\begin{equation}
q(T) = q_{cool} \exp (-T / T_{cool})^2
\end{equation}
where $q_{cool} = 2$ and $T_{cool} = 5 \times 10^5$ K; 
we shall discuss the sensitivity of the flow solutions 
to these two adjustable parameters. 
When $T \gg T_{cool}$, as usually occurs,
the cooling dropout term is negligible 
and has no influence on the flow. 
We use Equation (4) mainly 
to locate where gas is 
cooling; physically, it could represent the formation 
of stars or dense clouds that orbit near their radius 
of formation.
Flows calculated with $q = 0$ and a minimum temperature 
$T_{min} = 10^4$ K are essentially identical 
to those using Equation (4); in this case 
gas that cools to $T_{min} = 10^4$ K 
freefalls into the center of the galaxy. 
While our results are insensitive to the precise 
mathematical form of Equation (4), 
in our 2D calculations we prefer 
to remove cooled gas with a step function:
\begin{equation}
q(T) = \left\{ \begin{tabular}{lll}
	0 & {\rm for} & $T > T_{cool}$\\
	5 & {\rm for} & $T < T_{cool}$
	    \end{tabular} \right.
\end{equation}
where $T_{cool} = 5 \times 10^4$ K. 
This step function with its lower $T_{cool}$ and sudden onset 
better illustrates the geometrical complexity of 
of the cold gas distribution in 2D flows.

For 1D and 2D models we use a modified version of the 
ZEUS-2D code (Stone \& Norman 1992).
In 1D there are 300 zones within $r_{max} = 500$ kpc that 
slowly increase in size from the smallest zone at 50 pc. 
To capture the complex small scale structures of out 
2D calculations, we use a cylindrical grid with 
600 zones along each axis 
out to $z_{max} = R_{max} = 190$ kpc; the central zone 
has dimensions $z_1 = R_1 = 100$ pc. 
2D solutions with $120 \times 120$ or 
$300 \times 300$ zones (and $z_1 = R_1 = 100$ pc) 
are qualitatively similar.

The ZEUS code has been widely tested 
and is known to accurately compute a variety 
of standard hydrodynamical problems 
(Stone \& Norman 1992).
Our detailed source term modifications of ZEUS 
(Equations 1 - 3)
have been checked by comparing our galactic flow models 
with identical calculations using the 
independent cooling flow 
hydrocode developed by Annibale D'Ercole in Bologna 
(e.g. D'Ercole \& Ciotti et al. 1998).
Finally, we have successfully checked the 2D ZEUS code against 
analytic expressions for nonsteady inflow in hypothetical 
galaxies of uniform density (Appendix of Brighenti \& 
Mathews 1996).

In the following we discuss a large number of 1D and 2D 
flows, both heated and unheated. 
Readers who wish to review a representative 
sample of each type of model 
may elect to skip \S 6.3 and 6.4. 
However, the 1D and 2D flows described in 
\S 6.3 and 6.4 are essential in establishing the generality 
of our conclusion that heating cannot substantially 
reduce the global cooling rate in the flows.

\section{COOLING FLOWS WITHOUT HEATING}

For comparison with heated models it is useful to first 
calculate traditional cooling flows without heating
or other complications, i.e. 
$H(r,t) = F_{cond} = F_{conv} = 0$.

\subsection{No Cooling Dropout, q = 0}

The left column of 
Figure 1 shows the gas density, temperature 
and velocity at $t_n = 13$ Gyrs 
for single phase $q = 0$ flows in 
both isolated and accreting galaxies. 
It is seen that the NGC 4472 data can be matched 
reasonably well with the isolated solution for 
$1 \lta r \lta 10$ kpc, but the temperature 
is low and has the wrong gradient,
i.e. $dT/dr < 0$. 
In $r \gta 10$ kpc the gas density is also too low in the 
isolated galaxy.
The outer density 
profiles are better fit with the accreting solution 
(dashed lines),
indicating that not all of the hot gas in massive 
E galaxies 
comes from stellar mass loss (Brighenti \& Mathews 1999). 
Of particular interest for the $q = 0$ accreting 
solution in Figure 1 is its agreement with the positive 
temperature gradient observed beyond about 10 kpc.
In all well-studied galactic cooling flows the 
gas temperature rises from a minimum at the origin 
to a maximum near $\sim (3 - 4)r_e$ 
where the temperature 
is about 30 percent larger, then slowly decreases
at larger radii beyond the optical image of the 
central galaxy (Brighenti \& Mathews 1997).
High gas temperatures at large radii are produced in 
the accretion shock when the gas first entered the massive 
dark halo of the galaxy group surrounding the luminous  
elliptical. 
The post-shock temperature of the accreting gas 
is close to the virial temperature 
of the entire galaxy group.
The positive 
temperature gradient within several $r_e$ is produced 
as the temperature approaches the somewhat smaller 
$\sim 1$ keV virial temperature of the central galaxy. 
This temperature profile is consistent with that of 
NGC 4472 shown in Figure 1.
Finally, we note that the temperature $T(r)$ plotted in 
Figure 1 and in subsequent figures is shown as a function 
of physical radius $r$, not the slightly different 
emission-averaged temperature 
along the line of sight at the projected radius. 
Although this latter temperature profile 
would be more appropriate 
to compare with the NGC 4472 observations, 
$T$ as a function of $r$ is necessary 
to facilitate comparisons with 1D heated cooling flows
(\S 6 below). 
 
The computed density $n(r)$ in $0.1 \lta r \lta 1$ kpc 
for both $q = 0$ solutions exceeds the observations; 
indeed, the hypothesis of distributed cooling (mass dropout) 
$q > 0$ was first devised to correct this discrepancy 
(e.g. Thomas 1986; 
Vedder, Trester \& Canizares 1988;
Sarazin \& Ashe 1989).
However, 
the most obvious and fatal problem with 
$q = 0$ models is that 
$4 \times 10^{10}$ $M_{\odot}$ of cooled gas has 
accumulated in the central computational zone by 
time $t_n$. 
As a consequence of this mass, 
the hot gas density inside 100 pc 
rises sharply as seen in Figure 1.
This central mass 
is $\sim 20$ times larger than the black hole mass 
observed in NGC 4472 
and causes the gas temperature to peak unrealistically 
in $r \lta 5$ kpc, unlike any known bright elliptical galaxy.

As a historical point, 
in earlier steady state galactic cooling flow models 
the global temperature profiles were sensitive to the outer 
boundary conditions. 
White and Chevalier (1984) required that the gas 
density vanish with the flow velocity at 
some large stagnation radius $r_{stag}$. 
This constraint causes the temperature to rise 
with radius and become infinite at $r_{stag}$, 
assuming the gas pressure there is finite.
This sensitivity of steady state flows to the outer boundary 
conditions was recognized by  
Vedder, Trester \& Canizares (1988) whose 
solutions -- unconstrained by the 
density at $r_{stag}$ -- reveal a wide variety of 
global temperature profiles.
They realized that steady state 
boundary conditions imposed at $r_{stag}$ are 
unrealistic since the flow time to 
the origin typically exceeds the Hubble time.

In summary, isolated flows without cooling dropout have 
too little density beyond $\sim 10$ kpc and have negative 
rather than positive gas temperature gradients.
Accreting flows are in good agreement with the 
extended hot gas density structure and the positive 
temperature gradient within several $r_e$.
But both $q = 0$ solutions fail 
because the central mass of cooled gas is too large.

\subsection{With Uniform Cooling Dropout, q = 1}

Also shown in Figure 1 are traditional cooling flows 
with distributed (multiphase) 
cooling dropout at every radius, $q = 1$
(e.g. Sarazin \& Ashe 1989). 
Both the density and temperature are 
reasonably well matched, except in the central 1-2 kpc.
The temperature derivative is positive 
for the accreting model  
in $5 \lta r \lta 50$ kpc in agreement with observations. 
The temperature is slightly lower than observed in NGC 4472, 
but this could be corrected by small 
adjustments in the parameters
of the calculation (e.g. Brighenti \& Mathews 1999).
The NGC 4472 observations are included in the Figures 
only for reference and we have not attempted to fit them 
exactly. 

The central density in both $q = 1$ solutions 
is still too peaked, but a 
flat density core could be produced by adding 
magnetic or cosmic ray pressures not included in Figure 1.
The mass that cools at the origin is only $2.1 \times 10^9$ M$_\odot$
and $2.9 \times 10^9$ M$_\odot$
for the isolated and accreting $q = 1$ flows respectively, 
but the total mass of cooled gas is very similar 
to the $q = 0$ flows. 
Additional details on models of this sort can be found 
in Brighenti \& Mathews (2000).

While the $q = 1$ dropout solutions are generally 
successful, the distributed dropout hypothesis has been 
criticized for several reasons: 
(1) there is as yet no XMM-Chandra spectral evidence
for cooling to $T = 0$, 
(2) linear perturbations do not result in localized cooling,
(3) large masses of 
colder, cooled gas have not been observed, and 
(4) continuous star formation requires
deviations from the Salpeter IMF.

\section{HEATED COOLING FLOWS}

\subsection{When Heating Locally Balances Cooling}

Some authors have noted that the 
observed mass of hot gas in large E galaxies 
is roughly equal to the total mass ejected from stars,
$M_{gas} \approx M_{*t} \alpha_*(t_n) t_n \approx 10^{10}$
$M_{\odot}$, and have concluded from this that 
there is no need for mass dropout as long as 
some hypothetical heating source
balances radiative losses at every radius.
In the left column of panels in Figure 2 we show idealized 
solutions of this sort without distributed 
cooling ($q = 0$) and assuming 
\begin{equation}
\rho H(r,t) = { \rho^2 \Lambda \over m_p^2}
\end{equation}
so that radiative cooling is effectively zero. 
In these solutions $u \approx 0$ 
and no gas cools at any radius. 
However,  
the gas density is very large 
at time $t_n = 13$ Gyrs and much too centrally peaked 
($\rho \propto \rho_*$) due to the absence of 
an inward flow.
As the mass of interstellar gas accumulates, 
a larger fraction of the pressure gradient required for 
virial equilibrium is taken by the density gradient 
so the temperature and its gradient are somewhat low.
We have also computed flows for which 
the dissipation of stellar mass loss
$\alpha \rho_* \varepsilon_{o,*}$ is the only heating, i.e. 
\begin{equation}
\rho H(r,t) = { \rho^2 \Lambda \over m_p^2}
- \alpha \rho_*
\left[ \varepsilon_{o,sn} - \varepsilon - {P \over \rho}
+ {u^2 \over 2} \right]
\end{equation}
where
$\varepsilon_{o,sn} = (3 k / 2 \mu m_p)
(\alpha_{sn} T_{sn})/\alpha_*$, 
with results very similar to those shown in Figure 2.
While this type of globally heated 
flow might seem to be the most perfect 
possible solution to the mass dropout problem,
it fails miserably to fit the observed data.

\subsection{A Dripping Flow with Distributed Heating}

Before we discuss results of a large
number of flow calculations with various assumed 
heating terms $H(r,t)$, it is useful to consider in detail
a specific but representative flow 
with nonzero heating.
For this purpose we adopt a Gaussian profile:
\begin{equation}
H(r,t) = G_h \exp[-(r/r_h)^2]
\end{equation}
with $G_h = 1$ erg gm$^{-1}$ s$^{-1}$ and 
$r_h = 5$ kpc. 
The heating is instantaneously applied whenever the 
gas velocity into the central zone is negative, 
otherwise $H(r,t) = 0$.
The heating feedback tends to inhibit 
gas from flowing into the center.
This heating scenario is clearly a simplification,
but if additional features were included -- time
delay for feedback, slower rise or decay time for $H(r,t)$, 
slower propagation of heating through the gas, 
etc. -- these details would obscure but not
significantly alter our main results.
As noted earlier, we do not attempt here to justify
$H(r,t)$ in terms of any particular physical model
for heating, our approach is purely phenomenological.

In the right column of 
Figure 2 we show the flow variables for both
isolated and accreting models at time $t_n = 13$
Gyrs as influenced by the Gaussian heating of
Equation (8). 
At earlier times, $t < t_n$, 
the flow velocity into the central zone is generally 
negative, so the heating
$H(r,t)$ is almost 
continuous. 
The gas density is globally lower than observed in 
NGC 4472, but the shape is reasonably correct, 
particularly for the accreting flow.
However, the gas temperature in the accretion flow is 
too large for $r \gta 5$ kpc due to the central heating
and the slope $dT/dr < 0$, is wrong.

The striking 
new feature in this Gaussian-heated flow 
is a nearly stationary ``drip''
wave at $r \approx 7$ kpc visible in Figure 2 
where the gas density
has a sharp maximum and the temperature drops 
suddenly to very low values. 
(This is why we do not attempt to 
plot the average temperature along the line of sight.)
This type of concentrated drip cooling is a standard
attribute of 1D spherically symmetric 
heated models.
The drip in the right column of panels 
in Figure 2 occurred because
the heated gas within 7 kpc is slowly flowing out, 
while the negative velocity in $r > 7$ kpc behaves
much like a normal cooling flow.
Just at $r \sim 7$ kpc 
the gas is locally compressed, 
${\bf \nabla} \cdot {\bf u} < 0$, generating 
runaway radiative cooling.

The global velocity field has produced a 
spontaneous nonlinear thermal
instability far from the center of the flow.
We note that ``drips'' were first discovered
in unheated cooling flows (Mathews 1997)
where the differential
(negative) flow velocity can occasionally
produce a local compression.
If cooled gas is not removed from the drip, 
the mass of locally cooled gas increases until, 
in 1D cooling flows, it goes into freefall and 
collapses supersonically 
toward the center of the flow, 
producing reflected shocks from the origin and 
altering
the global density and temperature profiles in 
very unrealistic ways. 
Our drips move slower or even become approximately
stationary since we remove 
cooled gas inside the drip by 
adopting $q > 0$ (Equation 4).

Drips may have been missed or misinterpreted in
previous discussions of heated cooling flows.
They are particularly easy to capture in Lagrangian
calculations (Mathews 1997), but can be missed 
in Eulerian calculations particularly if the
grid is too coarse. 
One-dimensional drip waves, such as those shown in Figure 2,
are clearly unphysical since they
are unstable both to Rayleigh-Taylor disruption
(${\bf \nabla}\rho$ opposes 
${\bf g} = -{\bf \nabla}\Phi$)
and to convective motions (negative entropy gradient 
inside the drip structure).
The true evolution of drips can be understood only with 
2D or 3D calculations, discussed below,
but the incidence of spatially distributed cooling,  
represented in 1D with a drip, 
as well as the total mass and approximate location 
of cooled gas, are determined
reasonably well from 1D calculations.

We find for the isolated 1D galaxy (right column of Figure 2)
that $4.0 \times 10^9$ $M_{\odot}$ 
has cooled for $r<100$ pc
and an additional 
$6.1 \times 10^9$ $M_{\odot}$ has cooled in 
$0.1 < r < 9.5$ kpc due to inward propagating drips during 
$t < 3$ Gyr.
Figure 3 shows the mass distribution 
of cooled gas $M_{cool}(r)$  
for the isolated heated flow at two times, 
3 and 13 Gyrs, 
assuming that the cooled gas remains at its radius 
of origin.
We cannot fully trust the astronomical
accuracy of our flow solutions 
during the first 1-2 Gyrs of the calculation ($t < 3$ Gyrs).
Since the stellar mass loss is particularly copious 
at these early times, it is possible that relatively 
large masses of cooled gas combined and formed 
stars with a normal IMF.
By the current time $t_n$ 
these stars would be almost as old at the original 
population and would contribute to the luminous stellar 
mass within several kpc.
So the large mass
$4.0 \times 10^9$ $M_{\odot}$ that flowed
into the central grid zone before $t = 3$ Gyrs 
in this calculation, 
which is larger than any known galactic black hole, 
cannot with confidence be used to disqualify this model.
For this reason we distinguish in Figure 3 
the mass that cooled before and after $t = 3$ Gyrs.

After time $t = 3$ Gyrs almost all of the 
cooling gas in the accretion or isolated flows is 
formed in the nearly stationary drip at 7 kpc. 
The concentration of cooling gas at this radius is an 
artifact of spherical symmetry; in an otherwise identical 
2D flow (discussed below) 
we find more distributed cooling in $r < 10$ kpc, 
but the total mass of cooled gas is unchanged.
The cooling 
rate during the last 3 Gyrs is $\sim 1$ $M_{\odot}$ yr$^{-1}$.
This nearly continuous drip cooling 
would produce a multiphase spectrum in apparent 
disagreement with X-ray spectra observed with XMM 
in M87 and other similar large E galaxies. 
{\it Ironically, therefore, the heating process that 
has been invoked to eliminate distributed cooling in 
cooling flows has instead stimulated widespread cooling 
just beyond the region of intense central heating.}

Another undesirable 
characteristic of heated cooling flows is the 
negative temperature gradient within several tens of 
kpc from the galactic center, as indicated  
in the right column of Figure 2 for $r \gta 5$ kpc.
This variation is opposite to the thermal profile 
observed in all luminous, 
group-dominant elliptical galaxies.

Thermal conduction is likely to be important within 
the sharp 1D drip structure where the temperature gradients 
become large. 
Convection must also occur since the entropy gradient 
is locally negative.
Both processes are expected to broaden the drip structure. 
But the question of most interest is whether the
cooling dropout can be slowed or completely
stopped by including conduction or convection.
We show in the Appendix that the 
mass dropout is not significantly reduced 
by either conduction or convection.

\subsubsection{The Same Flow in Two Dimensions}

Spherically symmetric 1D drip waves are inherently unrealistic 
because of local density and entropy inversions.
We now describe the more realistic 
two dimensional evolution of the same 
Gaussian heated flow for the isolated galaxy 
assuming ${\bf F}_{cond} = {\bf F}_{conv} = 0$.
We use cylindrical coordinates 
since this geometry does not favor the spherical character of 
the initial drip evolution, i.e. 
if distributed drip cooling occurs in 2D cylindrical flow, 
it would certainly be expected in 2D spherical coordinates.
Figure 4 shows the density and velocity structure 
at $t_n$ for the 2D version of the isolated galaxy 
flow in the right column of Figure 2 
heated as described by Equation (8) 
with $G_h = 1$ erg gm$^{-1}$ s$^{-1}$ and $r_h = 5$ kpc.
Since the flow is heated from below, 
subsonic turbulence is generated
near the center, $r \lta 25$ kpc.

The most interesting result of the 2D 
flow is that strong spatially distributed 
mass dropout still occurs. 
Also shown in Figure 4 are contours of the 
cooled gas density generated 
with the step function $q(T)$ in Equation (5) with 
$T_{cool} = 5 \times 10^4$ K, 
again assuming that the cooled gas remains where it first 
appears in the flow. 
Because of Rayleigh-Taylor and convective instabilities,
the distribution of cooled gas is no longer confined to 
a narrow drip region as in the corresponding 1D flow (Figure 2),
but cooling still occurs in about the same region.
The tendency for additional cooling to occur 
along the $R$ and $z$ axes is a numerical effect: 
when convective eddies flow toward the 
$R = 0$ or $z = 0$ boundaries, the 
reflection boundary conditions result in local compressions 
and subsequent cooling.
However, widespread cooling clearly 
occurs throughout the 2D flow within radius 
$r \lta 10$ kpc when convective eddies 
encounter each other, producing local 
nonlinear compressions. 
(A similar localized compressive cooling 
far from the galactic center 
also occurred in the calculations of gas flows in 
S0 galaxies
by D'Ercole \& Ciotti 1998).
By time $t_n$ the total amount of cooled gas 
$M_{cool} = 3.8 \times 10^{10}$ $M_{\odot}$ 
is comparable 
with that found in the corresponding 1D flow. 
Azimuthally averaged hot and cooled gas densities, 
$n(r)$ and $n_{cool}(r)$, for
this 2D flow are also shown in Figure 4.

The maximum radial extent 
of the cooled gas in 2D flows increases as the critical 
dropout temperature in Equation (5) $T_{cool}$ is increased 
above $5 \times 10^4$ K. 
This can be understood in terms of our 1D solutions in which 
drips typically begin at larger
radii and slowly migrate inward as their amplitude
and mass increase. 
The temperature minimum in the drip structure 
deepens as the drip moves inward.
Consequently, 
increasing $T_{cool}$ in 2D flows systematically 
removes gas at somewhat larger radii.
Because of this sensitivity to the parameter $T_{cool}$ 
and to the grid resolution, 
we cannot be confident of the 
precise distribution of cooled gas shown 
in Figure 4, but the 
reality of spatially distributed cooling (dropout) 
in these heated flows is robust. 
The Gaussian heated cooling flow shown in Figure 4 
would therefore have a multiphase spectrum 
(and temperature profile) unlike those observed with XMM. 

In view of the overall similarity of 
1D and 2D calculations of heated cooling flows 
such as those in Figures 2 and 4, 
we assume that if cooling drips occur in 
1D, distributed cooling 
will also appear in a more realistic 2D flows 
having the same parameters.  
We have not proven this assumption, but it seems to
hold true for the cases we have checked. 
As a result, we can explore a wide variety
of heated cooling flows in 1D 
and assume that a comparable 
mass dropout would also appear in similar
2D calculations. 
The 1D heated flows are not only quicker to compute, 
they show more clearly where and why 
the initial cooling occurs; the corresponding 2D 
flows reveal more realistically the 
geometrical complexities of the subsequent cooling.

To illustrate the similarity of global cooling 
rates in similar 1D and 2D flows, 
we compare in Figure 5 
the total mass of cooled gas as a function 
of time $M_{cool}(t)$ for the 1D 
and 2D Gaussian heated models.
The agreement is very good. 
It is also remarkable that the total mass of gas 
that cools and the mean cooling rate during the last 
few Gyrs are insensitive to the presence 
of conduction or convection (see Appendix), 
the cooling dropout threshold $T_{cool}$, 
and the computational geometry, 1D or 2D.
In 2D the distributed cooling is intermittent, 
occurring in bursts 
alternating with quiescent periods of 
several Gyrs without dropout. 

Cooling intermittency in 2D calculations 
can be explained by a quasi-cyclical 
variation of the density profile. 
After each cooling episode, the hot gas density is lowered 
and turbulence-driven compressions are insufficient 
to produce local thermal instabilities. 
Later, after $1-2$ Gyrs when the hot gas density 
has increased due to 
stellar mass loss and inflow, post-compression radiative losses 
become more efficient in triggering cooling dropout. 
In Figure 6 we 
show the evolution of the hot gas density profile 
during a cooling episode for the  
2D flow shown at time $t_n$ in Figure 4. 
During the cycle the ROSAT band X-ray luminosity 
varies from $L_x = 1.42 \times 10^{42}$ ergs s$^{-1}$
at time $t = 8$ Gyrs to a maximum of 
$L_x = 2.66 \times 10^{42}$ ergs s$^{-1}$
at $t = 9.8$ Gyrs and then back to 
$L_x = 8.76 \times 10^{40}$ ergs s$^{-1}$
at $t = 11$ Gyrs. 
Unfortunately, this variation of $L_x$ by about 30 
is unlikely to be the source of the large scatter 
observed in the ($L_x$,$L_B$)-plane since cooling is 
not supported by XMM observations and the large 
excursions in the density 
(and X-ray surface brightness) profile shown in Figure 6 
are not typically observed 
(e.g. Trinchieri, Fabbiano \& Canizares 1986).

\subsection{More Flows with Distributed Heating}

In addition to the representative heated cooling flow 
that we just described, 
we have computed a large number of additional flows 
for both isolated and accreting galaxies 
with various heating scenarios $H(r,t)$.
Most of these are 1D flows, but we assume that the 
corresponding 2D solutions are qualitatively similar.
Our conclusion is that none of these flows is acceptable. 
Heated cooling flows typically fail one or (usually) more 
critical tests:
(1) significant ongoing cooling occurs within $\sim 10-20$ kpc
which would be observed as a multiphase component 
in XMM spectra,
(2) the temperature gradient in 1-10 kpc is negative 
or peaked 
in heated accretion flows, not positive as observed, 
(3) the density at 5-20 kpc is often 
too low in heated flows due to the redistribution of 
heated gas.
While it is necessary and important to carefully document 
the shortcomings of each type of heated galactic flow, 
readers with a low tolerance for computational detail 
and who accept these general conclusions
may wish to skip to \S 7.


We consider a wide variety of heating profiles beginning with 
the diffuse 
Gaussian profile in Equation (8) but with different 
parameters $G_h$ and $r_h$. 
A more concentrated inverse square heating is described with 
\begin{equation}
H(r,t) = C_h T_C r^{-2}
\end{equation}
where the time dependence arises because heating is 
turned off if there is no mass flow into the central grid zone. 
This one-parameter ($C_hT_C$) heating function is designed to 
simulate inverse Compton heating if 
$T_C \gg T_{vir} \sim 10^7$ K, but 
it can represent any form of unspecified 
concentrated heating in the galactic core.
To simulate Compton heating when $T_C$ is closer to $T_{vir}$,
we have also performed calculations with 
\begin{equation}
H(r,t) = C_h r^{-2} (T_C - T).
\end{equation}
In this case 
$C_h = (4 \sigma_T /3 m_e c^2) (3 k/2 \mu m_p) (L_{agn}/4 \pi)$
erg cm$^{2}$ g$^{-1}$ K$^{-1}$ s$^{-1}$ is an approximation 
to isotropic Compton heating using the Thompson crossection.

We also consider a heating term proportional 
to the stellar density. 
This type of heating was performed by simply increasing 
the current Type Ia supernova rate 
SNu$(t_n)$ above the value $0.06$ SNu that agrees 
with currently observed iron abundances 
(see Figure A1 in Boute 2000). 
This type of heating $H \rho \propto \rho_*(r)$ is continuous, 
independent of inflow into the central grid, but 
$H$ can also share the Type Ia time dependence, 
SNu $\propto t^{-s}$,
so that SNu$(t_n)$ and $s$ are the two adjustable 
parameters for this heating scenario. 
While solutions with $s = 1$ and SNu$(t_n) > 0.06$ 
clearly produce too much iron, 
our intention is to explore the possibility of some type 
of hypothetical star-related or extended 
heating possibly unrelated to supernovae.
Finally, we computed flows  
in which mass dropout 
is accompanied with 
simultaneous local heating feedback,
$\rho H(r,t)  \propto 
{\dot \rho} = q(r) \rho / t_{cool}$.


\subsubsection{Additional Gaussian Heated Profiles}

In the first two columns of Figure 7 we illustrate 
Gaussian-heated 
flows with $r_h = 5$ kpc as before but with
larger and smaller heating coefficients $G_h$. 
Solid and dashed lines represent isolated and accretion flows 
respectively.
For $G_h = 0.5$, half as large as that 
in Figure 2 (right column), 
the heating is insufficient to induce significant 
large scale drip cooling so 
most of the gas cools within the central 100 pc.
For the isolated flow this mass is 
$M_{cool}(r<100~{\rm pc})$, is $1.0 \times 10^{10}$ $M_{\odot}$
and $3.0 \times 10^{10}$ $M_{\odot}$ at 3 and 13 Gyrs and 
for the accretion model the masses cooled within 100 pc are
$0.74 \times 10^{10}$ $M_{\odot}$ at 3 Gyrs and 
$3.9 \times 10^{10}$ $M_{\odot}$ at 13 Gyrs. 
These high central masses are clearly unsatisfactory.
For the isolated flow
the hot gas density is generally too low and 
for the accretion flow the gas density gradient is 
too steep, although the density 
profiles are somewhat time-variable. 
For both flows the gas temperature becomes large 
near the center due to the cooled mass there; 
as a result $dT/dr$ is negative 
within 5 or 10 kpc, not positive as observed. 

For $G_h = 2$, twice as large as that in Figure 2 
(right column), 
the larger heating causes the gas density to become 
much lower than the NGC 4472 density, 
but a large amount of gas still cools 
within 100 pc in both isolated and accretion 
flows, about $2 \times 10^9$ $M_{\odot}$, comparable 
to the black hole mass in NGC 4472.
The gas temperature in the inner galaxy $r \lta 10$ kpc 
is far larger than observed and $dT/dr < 0$ for $r \gta 5$ kpc.
Continuous drip cooling near 10-12 kpc 
would produce observable multiphase spectra 
for either accretion or isolated flows. 
In a more realistic 2D version of these $G_h = 2$ 
calculations we 
expect that this cooling would be distributed within 
$\sim 15$ kpc, not concentrated in the 1D drip as shown in 
Figure 7 (second column).
Clearly, the $G_h$ parameter must be fine-tuned even to 
achieve a poor fit to the observations.

Next we vary the spatial scale $r_h$ of the heating, then 
increase $G_h$ until the heating begins to affect the flow. 
The accretion and isolated flows have similar 
problems as before. 
For $G_h = 0.5$ and $r_h = 20$ kpc 
(third column of Figure 7), subsonic outflow occurs 
at $r > 1$ kpc, but the gas interior to this radius flows 
to the origin most of the time, 
and the central mass is still much too large, 
$6 - 10 \times 10^{9}$ $M_{\odot}$. 
Drip cooling occurs at $r \sim 60 - 100$ kpc, 
which would produce a multiphase spectrum.
The gas temperature is too high beyond 5 kpc for 
both the isolated and
the accretion flow. 
For both types of flow 
the density is too low at all radii $r \gta 1$ kpc. 
Increasing $G_h$ further lowers the density even more 
without appreciably reducing the central cooled mass.
As expected, values of $G_h$ less than $0.5$ 
(keeping $r_h = 20$ kpc) 
result in strong central cooling and 
even larger central masses.

When the Gaussian heating is more concentrated,
$r_h = 1$ kpc then $G_h$ must be $20$ to influence the 
flow (right column of Figure 7).
For these heating parameters, a central subsonic
outflow is present at $t_n$ within $\sim r_h$, 
resulting in sustained drip cooling at 
$r \sim 1.5 - 2$ kpc.
Interior to the drip the gas temperature is much 
too high, $T \sim 2 - 4$ keV.
This same pattern occurs in both 
accretion and isolated flows.
While $3.7 \times 10^{10}$ $M_{\odot}$ cools 
in $1.5 \lta r \lta 2$ kpc, 
very little gas cools within 100 pc in these models.
The central density is flattened with 
a rather good fit to NGC 4472 
at all relevant radii $r \lta 10$ kpc, 
but is about three times too large for 
$2 \lta r \lta 6$ kpc in both accretion and isolated flows.
Overall, models with more centrally concentrated heating 
are in better agreement with observations on large scales, 
but the deposition of extremely large 
masses of cooled gas within a few kpc 
and the associated multiphase XMM spectra are 
the same undesired consequences encountered 
in other flows with hypothetical heating. 

\subsubsection{Concentrated Heating with $H \propto r^{-2}$}

In this series of flows 
the heating is described by Equation (9). 
For either isolated or accretion 
flows with $C_hT_C = 1.6 \times 10^{43}$ 
erg cm$^2$ g$^{-1}$ s$^{-1}$
(where for example $L_{agn} \approx 10^{46}$ erg s$^{-1}$
and $T_C = 10^8$ K) 
the density gradient at time $t_n$ is too steep  
(as shown in the left column of Figure 8).
About $4 \times 10^{10}$ $M_{\odot}$ cools in 
$r \lta 0.5$ kpc, which results in a high temperature core 
within $1$ kpc that is rarely if ever observed. 
If $C_hT_C$ is increased to $1.6 \times 10^{44}$ 
to reduce the mass that cools near the origin, 
the central gas density is 
too low and the flat density core is unacceptably large
(right column of Figure 8). 
The central temperature is also too high 
and multiphase cooling is expected at about 1.5-3 kpc.
As before, results are similar for either accretion
or isolated models.
If thermal conduction is added to distribute the 
thermal energy to larger radii, neither the density 
or temperature profiles are substantially improved
and drip cooling still occurs but at a somewhat larger radius.
In summary, 
as the cooling parameter $C_hT_C$ increases by a factor 
of 10, the solutions shift from ones in which too much 
gas accumulates near the origin to flows in which the 
gas density is unrealistically lowered by strong heating.
It is possible that an intermediate value of $C_hT_C$ 
would produce a satisfactory density profile, but the
temperature would still be too high and multiphase gas 
near the center seems unavoidable. 
The fine-tuning required, moreover, 
seems very implausible.

\subsubsection{Distributed Heating with $H \rho \propto \rho_*$}

It has often been suggested that some source 
of stellar heating, such as with Type Ia supernovae, 
could balance radiative losses in 
cooling flows, arresting the cooling process. 
Therefore, we briefly consider 1D isolated and accretion 
flows in which $H(r) \rho \propto \rho_*(r)$. 
This can be accomplished by artificially increasing 
the supernova rate SNu$(t) =~$SNu$(t_n)(t/t_n)^{-s}$ 
by increasing SNu$(t_n)$ or adjusting 
its time dependence. 
We describe several solutions 
for isolated and accreting flows in which SNu$(t_n)$ is increased
while keeping $s = 1$.
The transition from cooling concentrated near the 
galactic center to galactic winds is very sensitive to SNu$(t_n)$.
Three isolated solutions are shown in the left column 
of Figure 9. 
For SNu$(t_n) = 0.20$ there is a cooling inflow and 
$M_{cool} = 2.3 \times 10^{10}$ $M_{\odot}$ cools 
within the central 100 pc; 
with SNu$(t_n) = 0.23$ there is a partial outflow 
with cooling dropout near 3 kpc;
with SNu$(t_n) = 0.25$ there is a complete wind 
and no gas cools at any radius.
The accreting flows shown in the right hand column 
of Figure 9 exhibit the same sensitivity to SNu$(t_n)$
with inflow at SNu$(t_n) = 0.25$, partial outflow at SNu$(t_n) = 0.28$ 
and a full wind at SNu$(t_n) = 0.30$. 
For either type of flow, the temperature gradient 
is negative everywhere and the hot gas density 
suddenly drops to unacceptably low values as 
SNu$(t_n)$ exceeds a sharp threshold 
for wind solutions. 
In addition to its inability to reproduce the 
observed temperature and density profiles,
this type of heating with $H(r) \propto \rho_*$ 
cannot be due to Type Ia supernova -- 
if it were, the iron  
abundance $z_{Fe}$ would be extremely high; 
for example in the accreting  
SNu$(t_n) = 0.25$ solution $z_{Fe}$ would be 
more than 6 times solar within about 60 kpc.

\subsection{Additional 2D Heated Cooling Flows}

While the initiation of 
nonlinear thermal instabilities that appear in heated
cooling flows are clearly revealed in the 1D flows, 
the later evolution of these instabilities 
is inherently multidimensional because of
Rayleigh-Taylor and convective instabilities.
For this reason we discuss 2D versions of several 
heated cooling flows already described 
with spherically symmetric models. 
For simplicity, we consider only 2D isolated flows. 
To implement the feedback in 2D, heating is applied only 
when the mass flux at radius $r = 300$ pc is negative, otherwise 
we set $H = 0$ (models with continuous heating give
qualitatively similar results).
Cooled gas is removed using the step function procedure 
(Equation 5) with $T_{cool} = 5 \times 10^4$ K.

The left column of panels in 
Figure 10 shows the 2D version of the 
Gaussian-heated flow with $G_h = 20$ and $r_h = 1$ kpc 
corresponding to the solid line 1D solution in the 
right column of Figure 7.
The upper panel shows the hot gas density and velocity 
and the second panel 
shows contours of the density of 
cooled gas, both at time $t_n$.
The lower two panels show the azimuthally averaged 
hot and cooled gas densities respectively.
While most of the cooling along the vertical 
($z$) axis is 
a numerical artifact, as explained earlier, 
there is genuine small-scale 
cooling away from both axes at $r \sim 1 - 2$ kpc.  
This cooling is visible as a peak  
in the azimuthally averaged cooled gas density 
at $r \sim 1.5$ kpc in the bottom panel. 
Cooling in this region is expected from the 
1D isolated flow illustrated in the right column 
of Figure 7. 
The total mass of cooled gas is 
$3.0 \times 10^{10}$ $M_{\odot}$.
The azimuthally averaged hot gas 
density distribution at time $t_n$ 
shown in Figure 10  
is much too steep relative to NGC 4472 
and is steeper than the corresponding 
density profile in Figure 7.
We believe that this latter difference 
can be explained by the quasi-cyclic 
variation in the density profile associated with episodic 
cooling as shown in Figure 6; 
i.e. the 1D and 2D flows are captured at different
phases of this cycle at time $t_n$.
Like the corresponding 1D flow, 
the 2D gas temperature (not shown) is also somewhat
too high in the central kpc, $T \sim 1.5$ keV.
This 2D heated cooling flow suffers from 
the usual astronomical problems: 
the mass of cooled gas within a few kpc is very 
large, radiatively cooling gas in this region should 
produce a multiphase XMM spectrum, and the 
variable excursions in 
the hot gas density profile are larger than 
the narrow range of density profiles 
consistent with X-ray observations 
(e.g. Fabbiano, Trichieri \& Canizares 1986).

The central and right hand columns of Figure 10  
show 2D flows at time $t_n$ corresponding to the 
isolated 1D flows illustrated in Figure 8 for which the 
feedback heating is Compton-like, $H \propto r^{-2}$.
The 2D solutions confirm our previous conclusions 
for these flows: 
for weak heating, $C_h T_C = 1.6 \times 10^{43}$, all the gas
($\sim 4 \times 10^{10}$ $M_{\odot}$) cools at the 
center of the flow (central column Figure 10) and the
gas density profile is too steep.
If the heating is stronger,
$C_h T_C = 1.6 \times 10^{44}$,
(right column Figure 10) widely 
distributed cooling occurs that is not seen in XMM spectra. 
Even with this larger $C_hT_C$, an excessively 
large mass $2.2 \times 10^{10}$ $M_{\odot}$ 
still cools throughout the galaxy.
The central gas density is unrealistically 
low within about 2 kpc and too high beyond.
The temperature is too high ($T \sim 1.5-2 \times 10^7$ K)
inside $\sim 10$ kpc, similar to the 1D solutions 
in Figure 8. 
Extended periods during which there is very little 
radiative cooling at large $r$ are punctuated 
by short episodes of intense cooling, 
but the average 
cooling rate history is similar to the other models shown 
in Figure 5.
Nevertheless, 
it is obvious that neither of these 
models with $H \propto r^{-2}$ is acceptable, 
nor is it plausible that any value of $C_hT_C$ 
could give acceptable results.

Finally, we have computed 1D and 2D versions of cooling 
flows in which the cooled
gas is rapidly heated at the cooling site, 
as if by Type II supernovae. 
In this idealized immediate feedback heating scenario 
$\rho H(r,t)$ is proportional to $q(r)\rho/t_{cool}$ so 
$q$ is no longer merely a means of removing cooled gas, 
but is directly related to the heating. 
In the 2D version of this heating model, in which $q$ is 
a step function described by Equation (5), 
we find that the convective turbulence is 
sufficiently strong to generate spontaneous 
and widely distributed radiative cooling within 10 - 20 kpc.
The azimuthally averaged gas density shows a cyclical
variation similar to that described in section 6.2.1.
Again, the central temperature is generally too high.

When $q$ is assumed constant (e.g. $q=1$), these models 
have some positive features,
such as naturally producing the observed core 
in the hot gas density.
However, they cannot be accepted since multiphase X-ray spectra 
would be expected from the cooling gas and
the central masses often exceed several $10^9$ 
$M_{\odot}$. 
Type II supernovae have never been observed 
in early type galaxies, so the physical origin 
of the heating process is obscure.

In summary,
by our rigorous standards for success -- no multiphase gas,
agreement with observed profiles, no large
masses cooling close to the center, etc. --
all of the flow models we have discussed,
as well as many others that we have calculated,
have failed to match the observations.

\section{Models with Non-thermal Pressure}

As we discussed in the Introduction, 
spatially extended non-linear thermal 
instabilities arise in cooling flows 
whenever the thermal gas is 
globally compressed, ${\bf \nabla} \cdot {\bf u} < 0$, 
or $\partial^2 (r^2u) / \partial^2 r < 0$ in 1D. 
For a variety of centrally heated flows we have shown 
that runaway radiative cooling is expected whenever 
the normally inflowing gas is decelerated 
and compressed by heated 
gas closer to center of the flow.
A similar deceleration can be produced 
in the absence of heating by an additional 
non-thermal pressure (gradient) which is dynamically 
important within some radius in the flow. 
The non-thermal pressure $P_{nt}$ could be due to cosmic rays 
or magnetic energy density perhaps 
associated with a central AGN.
For simplicity, and in keeping with
our purely phenomenological approach, 
we consider two pressurized flows in which the 
non-thermal pressure is assumed to scale with the thermal 
pressure $P$:
\begin{equation}
P_{nt} = P (r /r_{nt})^{-1}.
\end{equation}
The coefficient of $P$ is designed  
so that $dP_{nt}/dr$ is steeper than $dP /dr$.

The two columns in 
Figure 11 illustrate the flow patterns at time $t_n$ 
in isolated 1D galactic flows with $r_{nt} = 1$ and 2 kpc 
respectively.
With $r_{nt} = 1$ kpc the non-thermal pressure flattens the 
inner density profile somewhat like the NGC 4472 observations.
Because the central flow is partially supported by 
the additional non-thermal pressure, 
both the thermal pressure and temperature 
are also lower at small $r$.
But when $r_{nt} = 1$ kpc the non-thermal pressure gradient 
is insufficient to keep large masses of gas 
from cooling within 100 pc, 
$M_{cool}(r<100~{\rm pc},t_n) 
= 1.7 \times 10^{10}$ $M_{\odot}$.
However, if the non-thermal pressure is larger, 
$r_{nt} = 2$ kpc, gas slowly flows out from the 
core and a 1D drip forms at $r = 0.6$ kpc.
In this 1D flow all the gas cools at  
or near the drip, $r = 0.5 - 2.5$ kpc, not at the origin.
Note that $P_{nt}/P$ must be quite large in this model before 
gas is forced to cool away from the origin.
Nevertheless, neither central or 
extended cooling is supported by XMM observations.

\section{CONCLUSIONS}


For many years it has been assumed 
that cooling flows are inherently inhomogeneous 
(multiphase), so that thermal instabilities develop 
throughout extended regions of the flow. 
While there is little or no support for localized cooling 
from linear perturbation studies, 
by assuming distributed cooling dropout 
it has been possible to consume hot gas 
in a large volume that would otherwise have flowed to 
the very center and cooled there, generating undesirable 
central mass concentrations and unrealistically 
peaked X-ray emission and thermal temperatures.
If this multiphase hypothesis is correct, some 
spectral signature of gas cooling to low temperatures
is expected, such as a broadened FeL emission feature.

It is noteworthy therefore that 
the first XMM observations of elliptical galaxies and 
clusters have failed to find spectral 
evidence for locally multiphase flows, even in the bright 
cores of galaxies like M87 (Molendi \& Pizzolato 2001) 
and NGC 4636 (Xu et al. 2001).
These observations conflict with FUSE detections of OVI emission 
from gas at $T \sim 3 \times 10^5$ K  
observed in several bright elliptical galaxies.  
The OVI emission is 
a clear indication that gas is cooling, although
at a somewhat lower rate than expected
(Bregman, Miller \& Irwin 2001).
Nevertheless, the null XMM observations 
have revived interest in heating mechanisms 
that have been proposed to keep gas from cooling to 
low temperatures.
If some form of continuous or intermittent heating were present, 
it is claimed, our failure to observe cooling and cooled 
gas with XMM could be easily understood. 
The limited temperature variations that are observed 
at any projected radius could be 
explained simply by single-phase temperature variations 
viewed along the line of sight. 

To explore this suggestion in more detail 
we have computed a large 
number of cooling flows that are heated by feedback 
when gas flows into the center.  
This kind of feedback is ideally suited to keep gas from 
cooling near the center of the flow where the gas is 
densest and radiative losses are maximal. 
The heating scenarios we adopt 
are deliberately kept as simple as 
possible with a minimum number of free parameters, 
even though they are unphysical in some respects. 
Our approach is phenomenological, with 
little or no attention to specific 
(poorly understood) microphysical heating processes. 
We have also investigated how 
thermal conduction and convection can share this heating 
with larger galactic volumes.

The criteria that we impose for successfully heated 
models are exacting: 
(1) the density and temperature profiles, 
$n(r)$ and $T(r)$, at time $t_n = 13$ Gyrs must 
not differ greatly from
those observed in typical massive 
elliptical galaxies such as NGC 4472, 
(2) little or no gas must be currently cooling 
to low temperatures (i.e. $\ll 1$ keV) anywhere in the flow, 
${\dot M} \lta 0.1$ $M_{\odot}$ yr$^{-1}$ for 
NGC 4472, and
(3) the total mass of gas that has cooled near the 
galactic center in the last 
$\sim 10$ Gyrs must not exceed the total mass typically 
indicated by observed stellar velocity dispersions.

After examining a large number of heated cooling flows, 
much larger than we have described here,
we conclude that it is very difficult
to devise any combination of {\it ad hoc} heating, thermal
conduction or convection that produces reasonable density
and temperature profiles and keeps gas from cooling
at rates much less than 
that implied by the radiative emission of X-rays, 
$${\dot M} \approx \left( {2 \mu m_p \over 5 k T}\right) L_{x,bol} 
\approx 2.5~~M_{\odot}~{\rm yr}^{-1},$$
which is 
evaluated for the temperature and bolometric luminosity 
of NGC 4472, $T \approx 1.3 \times 10^7$ K 
and $L_{x,bol} \approx 7.2 \times 10^{41}$ erg s$^{-1}$.
This result is illustrated most clearly in Figure 5, 
showing that neither the total mass of cooled gas nor the
present cooling rate 
${\dot M}(t_n) = dM_{cool}/dt \approx 1.2$ $M_{\odot}$ yr$^{-1}$
is appreciably reduced 
in any of the heated cooling flows we consider, many of which 
are extreme in unphysical ways.
Even continuous heating that exactly balances radiative  
cooling fails spectacularly to agree with observations. 
The central temperatures in heated flows is typically
too large and the central temperature 
derivative $dT/dr$ is often negative, 
unlike all well-observed flows. 
As a consequence of these high temperatures, the gas densities 
are too low in the center and (often) throughout the 
entire flow. 
When heating lowers the gas density, we find that 
Type Ia supernova raise the hot gas iron abundance 
far beyond observed values; thus unrealistically high 
iron abundances are an additional serious difficulty with 
heated cooling flows that we have not emphasized here.
Even when the heating parameters are adjusted for the 
best (but nevertheless unacceptable) fit to the observations,
the solutions are found to be very sensitive to the 
heating parameters. 
An undesirable degree of  
fine-tuning appears to be essential for any 
flow influenced by heating. 

We conclude that the failure of XMM to detect cooling gas 
at the expected rate cannot be easily understood in terms of some 
as yet unspecified heating 
mechanism that balances or exceeds the 
X-ray luminosity. 

%
%
%
%

While the XMM results are inconsistent with multiphase 
flows that cool, optical (H$\alpha$ + [NII]) and 
UV line emission have been used to support 
radiative cooling.
We estimate from the FUSE flux for the OVI doublet 
in NGC 4636 (Bregman, Miller \& Irwin 2001)
that the OVII 21.6\AA (0.574 keV) 
line should be about 10 percent 
as luminous as the strong OVIII Ly$\alpha$ line 
detected by Xu et al. (2001) using XMM/RGS. 
Yet Xu et al. were unable to detect the 
OVII line expected 
from gas cooling near $T \sim 2 \times 10^6$ K 
and we have no explanation for this. 
Nevertheless, our conclusion here remains: 
if emission from low 
or intermediate temperature ions is absent, 
this is unlikely to be 
explained by some hypothetical source of 
heating from the galactic center.
If the heating is insufficient to drive a global galactic wind, 
our heated flow calculations show convincingly
that the heating actually stimulates gas to cool
in a large volume of the galaxy.
If the heating rate is large enough to drive a wind, 
the gas density (and $L_x$) will be too low everywhere, 
typically by several orders of magnitude.

On the positive side, we have demonstrated here how gas can 
cool many kpcs from the galactic center when there is 
some central heating or non-thermal pressure. 
Non-linear thermal instabilities develop naturally 
and spontaneously in the compressed region of the flow 
where the normal cooling inflow from 
large radii confronts slowed or outflowing gas 
in the heated or pressurized core. 
Strong local cooling occurs even when the flow velocities 
are subsonic.
The 1D representations of this instability, 
which we refer to as ``drips'', are 
unphysical since they are Rayleigh-Taylor 
and convectively unstable. 
If the cooled gas is removed,
spherically symmetric 1D drips usually become 
quasi-stationary at late stages of evolution. 
If the cooled gas is not removed, drips can 
evolve into massive, high density 
shells of cooled gas that ultimately  
freefall supersonically into the center of the flow, 
causing unphysical heating and shock reflection. 

It is gratifying therefore 
that cooling instabilities form at nearly the 
same radius in 1D and more realistic 2D flows.
In 2D flows random, subsonic 
(convective and turbulent) velocities 
induce cooling over a large galactic volume. 
When convective eddies collide, 
nonlinear compressions lead to runaway thermal 
instabilities.
The cooling is more chaotic in 2D and 
coherent inward-moving drip waves do not form. 
In principle, 
gas can also cool at intermediate 
radii in unheated galactic cooling flows 
provided the galaxy is rotating
(Brighenti \& Mathews 1997b, 2000b; 
D'Ercole,  \& Ciotti 1998), 
in this case an extended disk of cooling gas is formed, 
but there is no observational evidence for this 
in the X-ray images. 
We have shown that 
cooling disks can be avoided if the flow is 
turbulent or convective (Brighenti \& Mathews 1900) 
and there is considerable observational evidence 
for this (e.g. Caon et al. 2000).

Nonlinear cooling 
can also be stimulated by an additional
non-thermal pressure near the center of the cooling flow. 
This pressure could be due to magnetic field stresses,  
cosmic rays or both. 
Seed magnetic fields ejected from mass-losing stars 
can be amplified by the subsonic turbulence that results from 
Type Ia supernovae or by the random motions
observed in cooling flows (Caon et al. 2000). 
The field amplification due to this turbulent dynamo 
saturates when the magnetic energy density becomes comparable 
with that of the subsonic turbulence, 
which is generally less than the thermal energy density 
(pressure) in the hot gas. 
However, as the field is advected toward the center of the 
cooling flow, $B^2/8\pi$ grows by adiabatic compression 
and may exceed the gas pressure at the center
(Soker \& Sarazin 1990; Mathews \& Brighenti 1997). 
Therefore, a large non-thermal magnetic pressure 
could be a natural consequence of cooling flows even 
in the absence of AGN.
Alternatively, large non-thermal cosmic ray 
and magnetic pressures 
may be produced by AGN activity associated with 
central black holes.


\vskip.1in
\noindent
Our main conclusions are:

\vskip.1in
\noindent
(1) Cooling flows that are assumed to be centrally heated 
develop nonlinear thermal instabilities on spatial 
scales comparable to that of the hypothetical heating.
Thus heating from an active nucleus cannot 
explain the apparent absence of multiphase 
cooling gas in XMM spectra. 
This is particularly true if the heating is large enough 
to keep unrealistic masses of gas from cooling within 
the central kpc.

\vskip.1in
\noindent
(2) Central regions of high nonthermal pressure can 
also stimulate nonlinear thermal instabilities.

\vskip.1in
\noindent
(3) In most or all well-observed cooling flows 
in galaxy groups or clusters the 
gas temperature has a minimum at the center and 
increases to a maximum typically at several stellar effective 
radii in the group-central E galaxy. 
The higher temperatures in the more distant gas 
can be simply understood
in terms of the cosmic accretion of gas onto the
group dark halo which has a higher virial temperature
than the central galaxy.
As this gas flows inward, its temperature adapts to the 
lower virial temperature of the stars in the central
galaxy, resulting in a positive temperature gradient. 
None of our computed cooling flows with central heating 
has a monotonically increasing temperature 
profile as observed in this central region.

\vskip.1in
\noindent
(4) If the central heating is strong enough to keep 
unrealistically 
large masses of gas from cooling in the central few kpc,
the global hot gas density profile is usually radically
altered, and made time dependent. 

\vskip.1in
\noindent
(5) Flows heated by central feedback 
often result in eposidic cooling that
is associated with quasi-cyclic changes in the hot gas density
profile; at present there is no observational evidence for this.

\vskip.4in


\vskip.4in
We thank referee Peter Thomas for his thoughtful 
commentary on the first 
draft of this long paper and for his many useful suggestions.
Studies of the evolution of hot gas in elliptical galaxies
at UC Santa Cruz are supported by
NASA grant NAG 5-8049 and NSF grants  
AST-9802994 and AST-0098351 for which we are very grateful.
FB is supported in part by grants MURST-Cofin 00
and ASI-ARS99-74.


\clearpage
\vskip.1in
\figcaption[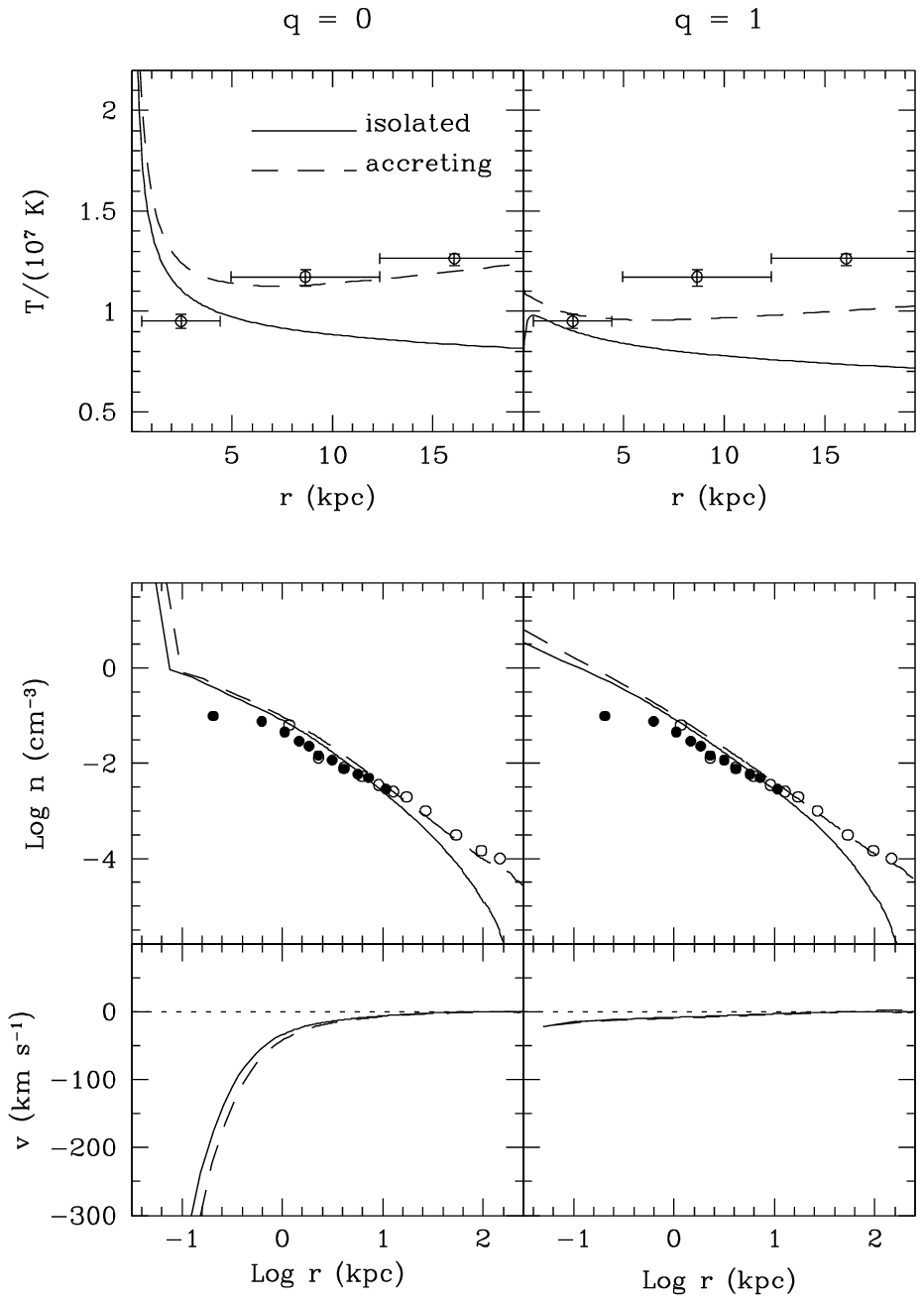]{
Gas temperature, density and radial velocity at 
time $t_n = 13$ Gyrs for 1D isolated 
({\it solid lines}) and accreting ({\it dashed lines}) 
galactic cooling flow models 
with two cooling dropout assumptions.
{\it Left column}: 
Flows with no distributed cooling dropout, $q = 0$.
{\it Right column}: Flows 
with uniformly distributed cooling, $q = 1$.
The temperatures are shown as functions of physical radius, not
averaged in projection along the line of sight.
Density and temperature observations for NGC 4472  
are {\it Einstein} data (Trinchieri, Fabbiano,
\& Canizares 1986) ({\it filled circles}) and 
ROSAT data (Irwin \& Sarazin 1996) ({\it open circles}).
\label{fig1}}

\vskip.1in
\figcaption[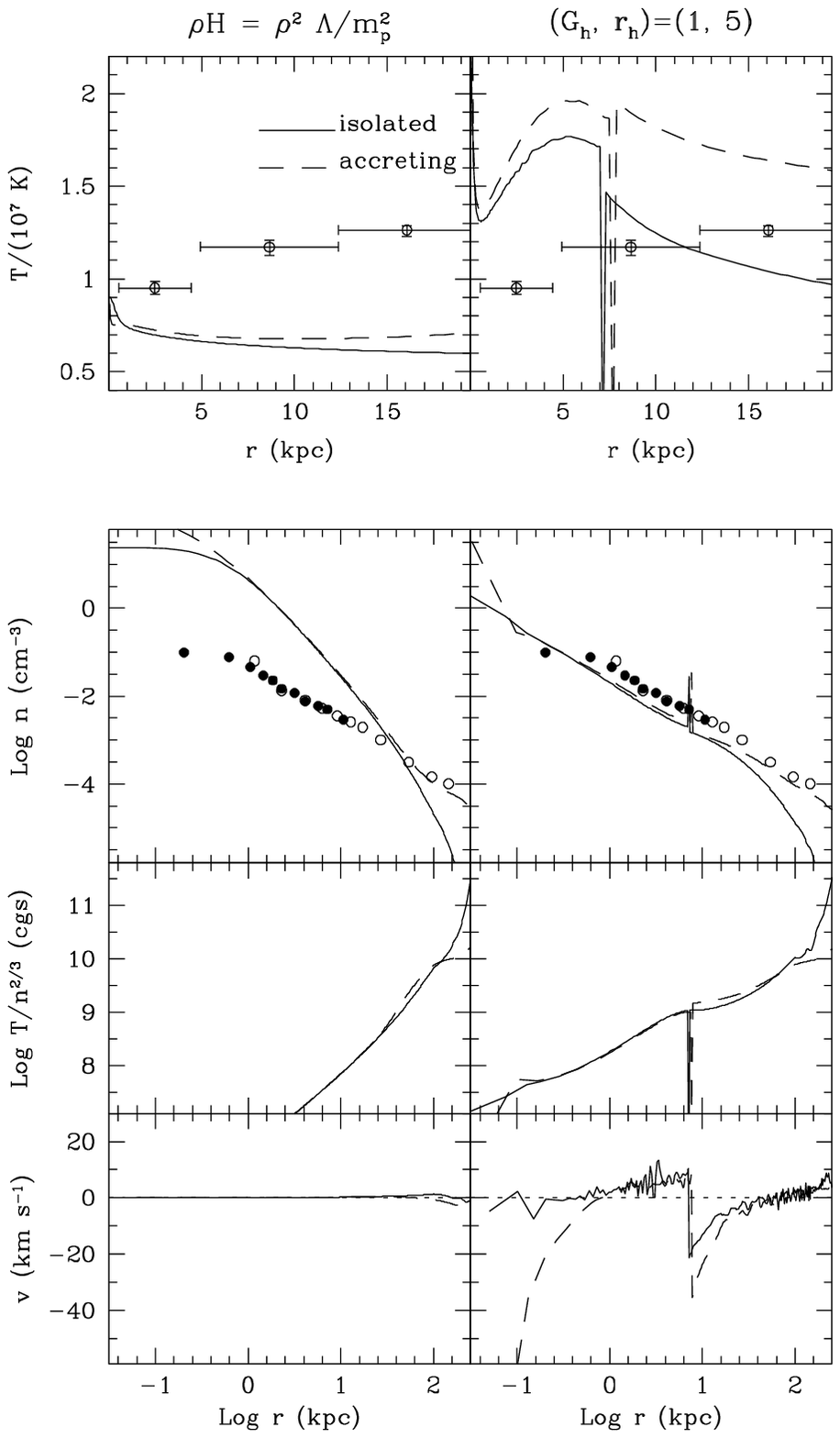]{
Gas temperature, density, specific entropy and 
velocity at 
time $t_n = 13$ Gyrs for four 1D heated 
galactic cooling flows. 
Each panel shows flows for isolated ({\it solid lines}) and 
accreting ({\it dashed lines}) galaxies.
The temperatures are shown as functions of physical radius, not 
averaged in projection along the line of sight.
{\it Left Column}: A galactic cooling flow model
in which cooling is locally balanced by heating.
{\it Right Column}: Comparison of flows 
heated as described in Eq. (8) with
$G_h = 1$ erg g$^{-1}$ s $^{-1}$ and $r_h = 5$ kpc.
A drip is seen at $r \approx 7-8$ kpc 
for both flows.
Data is the same as in Figure 1.
\label{fig2}}

\vskip.1in
\figcaption[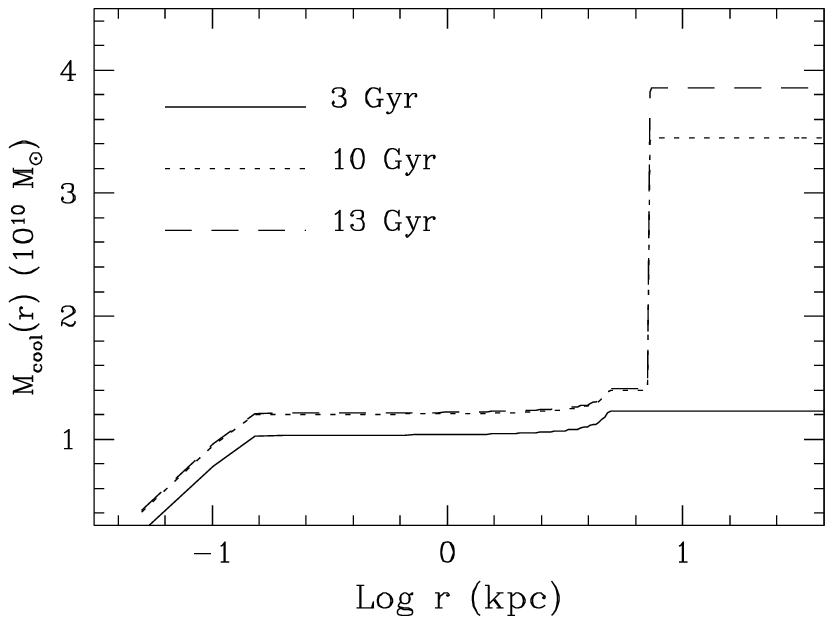]{
The integrated mass of cooled gas $M_{cool}(r)$ 
at various times for the heated cooling flow in the isolated 
E galaxy shown in the right column of Figure 2.
Note the $M_{cool}(r)$ plot is linear, not logarithmic.
\label{fig3}}

\vskip.1in
\figcaption[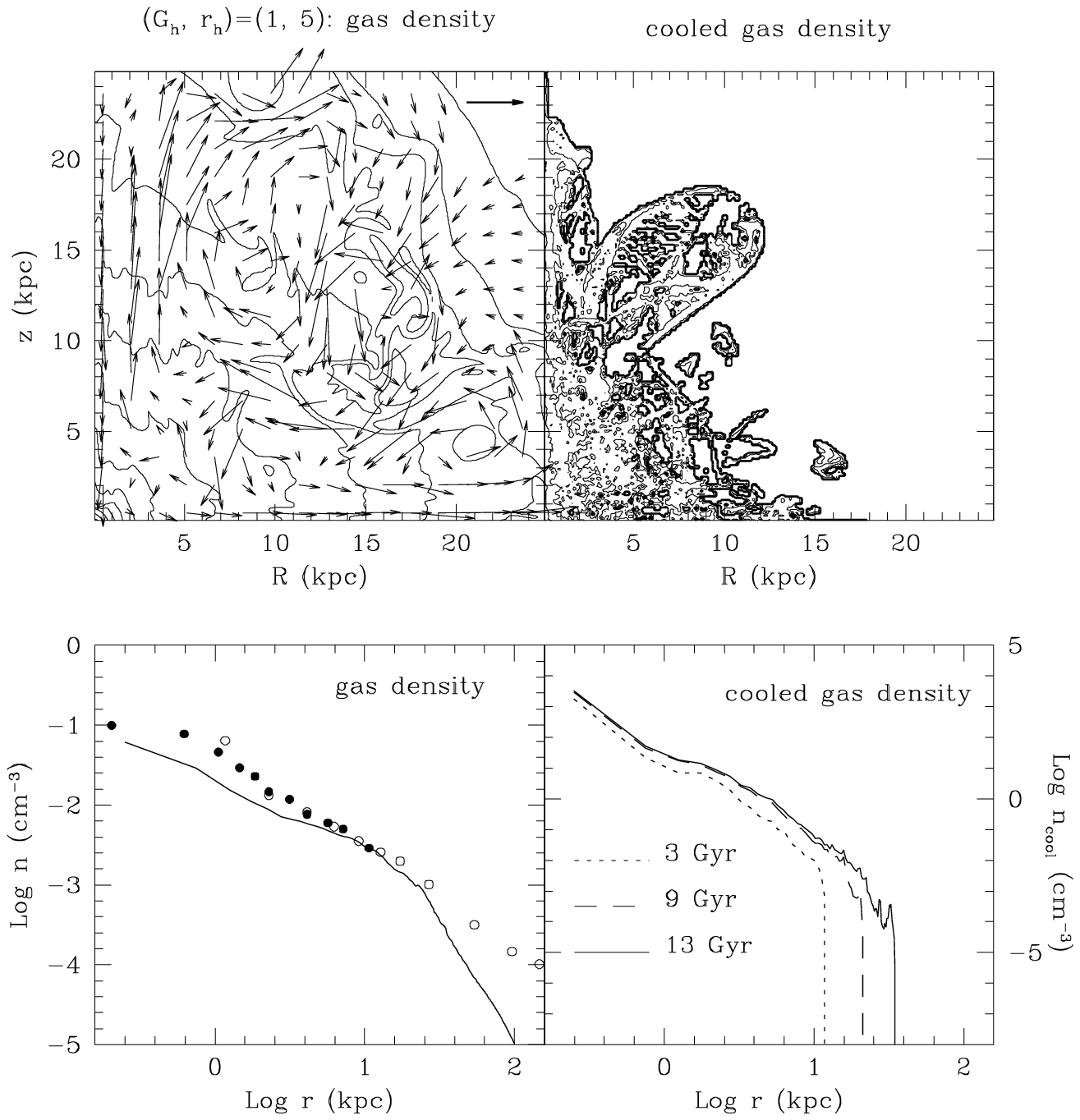]{
Four panels show the
2D flow with Gaussian heating ($G_h = 1$, $r_h = 5$ kpc)
at time $t_n = 13$ Gyrs.
{\it Upper Left}: Gas velocity and density. Contours
show $\log \rho(R,z)$. 
The velocity arrow at the upper right represents 
200 km s$^{-1}$.
{\it Upper Right}: Density contours of cooled gas. 
{\it Lower Left}: Azimuthally averaged hot gas density
shown with the same data as in Figure 1.
{\it Lower Right}: Azimuthally averaged cooled gas density
shown at three times.
\label{fig4}}

\vskip.1in
\figcaption[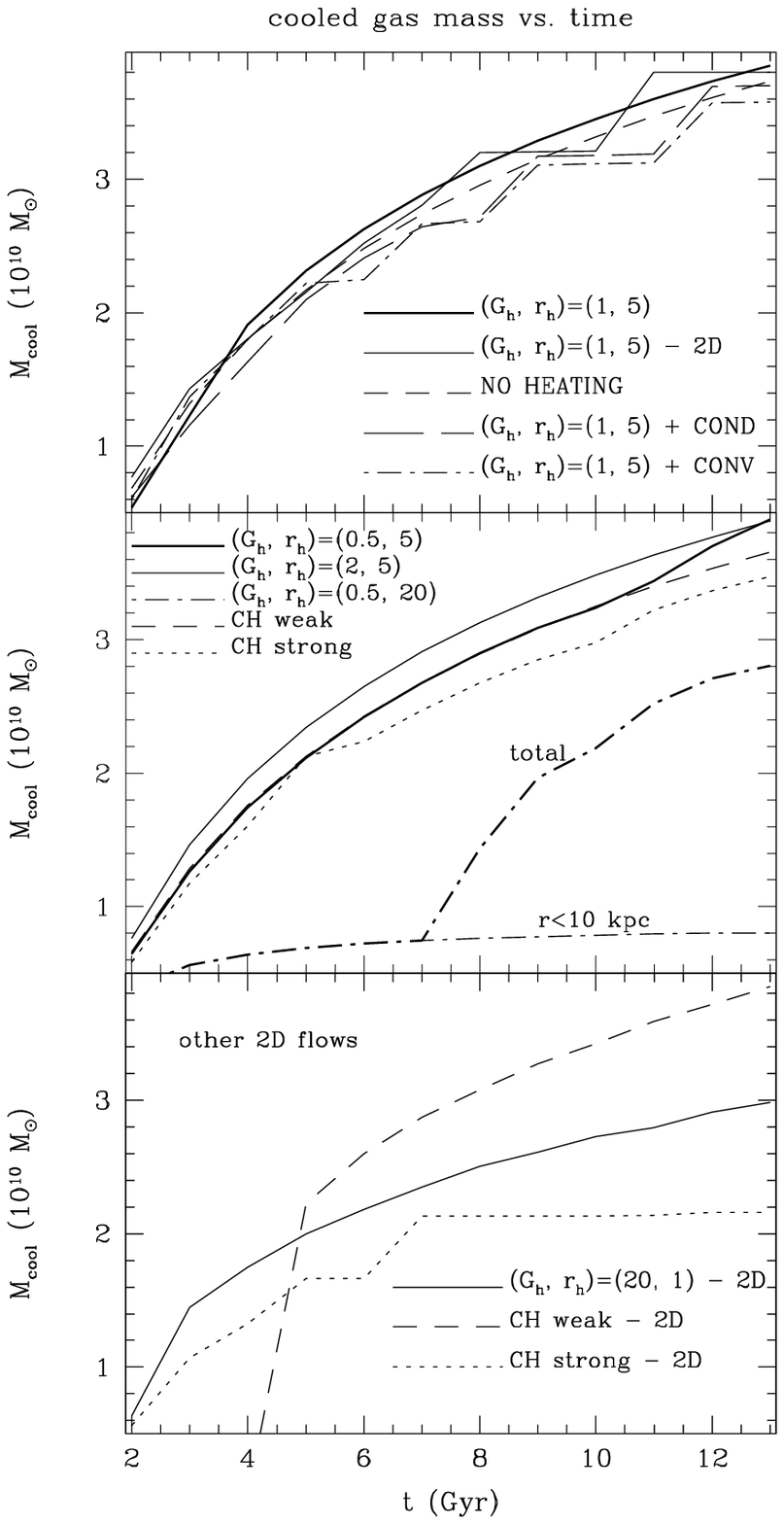]{
Plots of the growth of cooled gas mass $M_{cool}(t)$
for a large number of heated cooling flows. 
All flows are for the ``isolated'' galaxy. 
Flows with Gaussian heating are shown with the 
values of the two heating parameters in 
Equation (8), $G_h$ and $r_h$. 
Gaussian-heated flows computed in 2D are labeled; 
all other flows are in 1D including the two flows with 
conduction and convection.
Flows heated as described by Equation (9) are labeled CH; 
weak and strong heating refer respectively to 
$C_hT_C = 1.6 \times 10^{43}$ and $1.6 \times 10^{44}$ 
erg cm$^2$ g$^{-1}$ s$^{-1}$. 
{\it Top Panel}: $M_{cool}(t)$ for four Gaussian heated flows 
all with $G_h$ = 1 erg gm$^{-1}$ s$^{-1}$ 
and $r_h = 5~{\rm kpc})$ compared with a flow with no heating.
{\it Middle Panel}: $M_{cool}(t)$ for 
three Gaussian heated flows and 
two CH flows. For the ($G_h$,$r_h$)$ = (0.5,20)$
flow the cooled mass within 10 kpc is shown separately.
{\it Bottom Panel}: $M_{cool}(t)$ for three 2D heated flows. 
\label{fig5}}

\vskip.1in
\figcaption[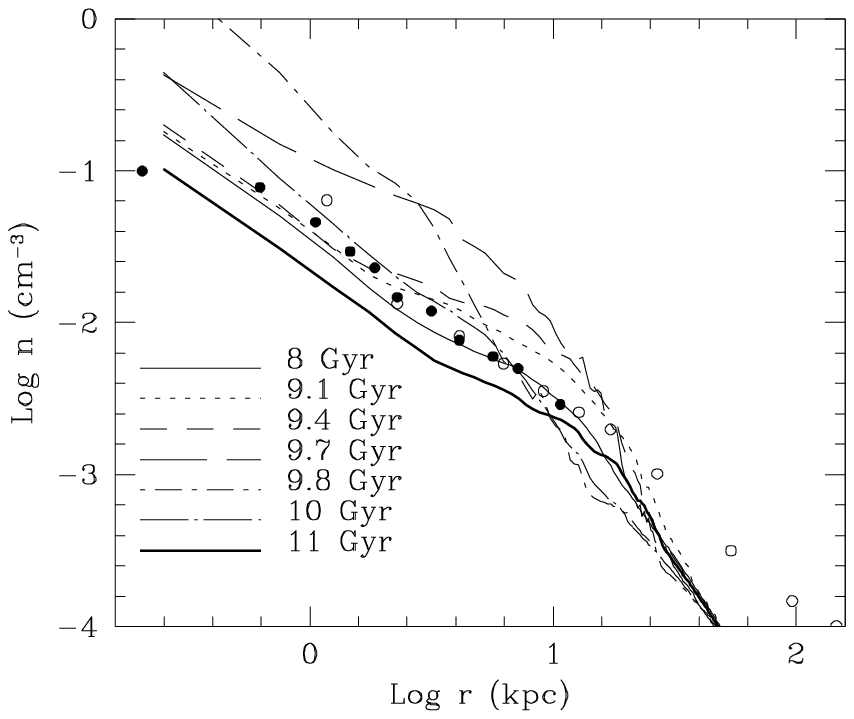]{
The density cycle during a cooling event. 
Azimuthally averaged hot gas density profiles 
for the 2D flow heated with ($G_h$, $r_h$) = (1, 5 kpc)
are shown at several times preceding, during and
following a cooling episode.
Times in Gyrs label each density profile.
The times and total masses of cooled gas 
are: ($t$ Gyrs; $M_{cool}/10^{10} M_{\odot}$)
= (8.0; 3.185), (9.1; 3.19), (9.4; 3.194),
(9.7; 3.283), (9.8; 3.618),
(10; 3.660) and (11; 3.685).
Data is the same as in Figure 1.
\label{fig6}}

\vskip.1in
\figcaption[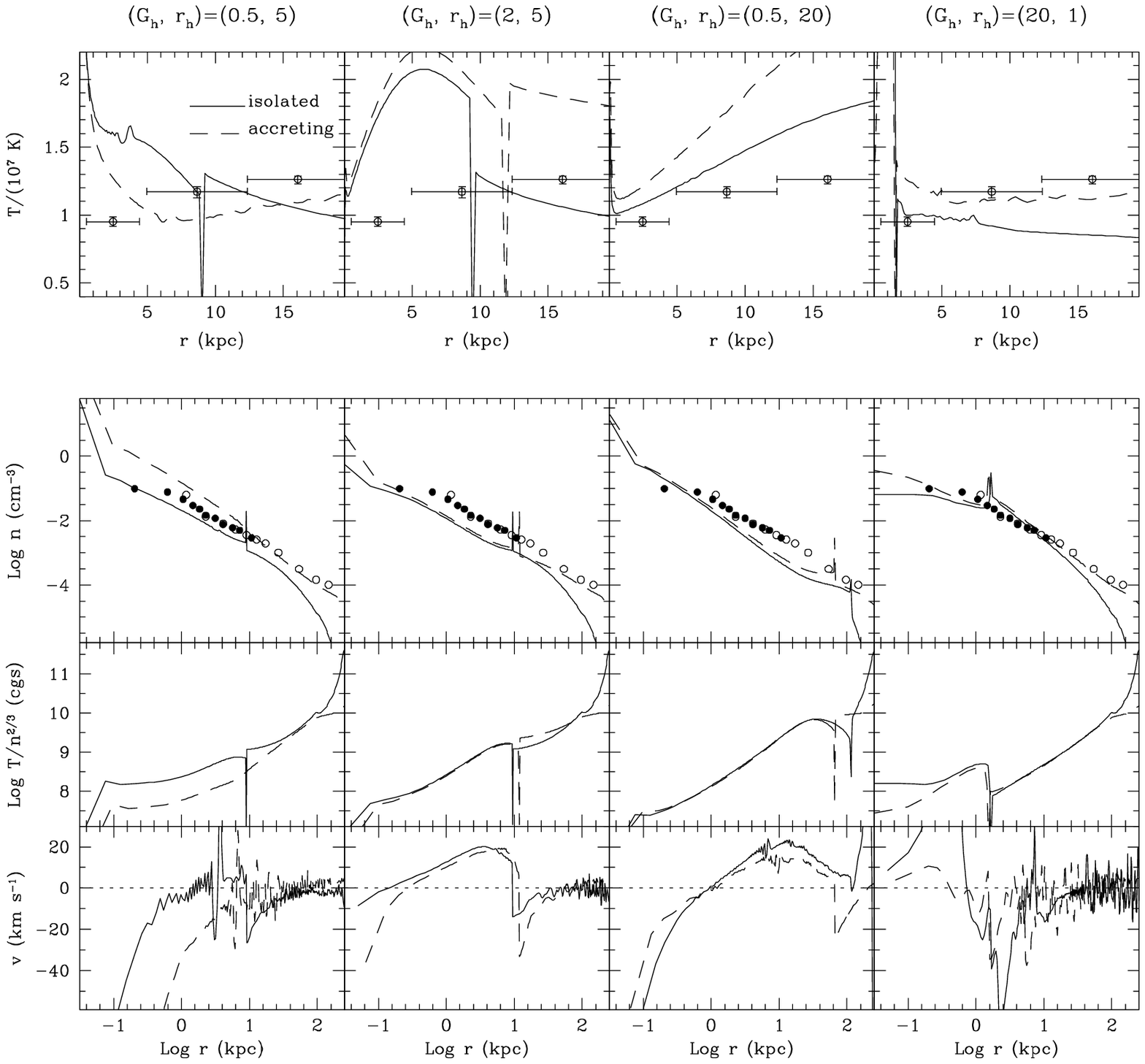]{
Gas temperature, density, specific entropy and
velocity at
time $t_n = 13$ Gyrs for four 
Gaussian heated cooling flow models. 
Flows for isolated and accreting
galaxies are shown with {\it solid lines} and
{\it dashed lines} respectively. 
Gas temperatures are shown as functions of physical radius, not
averaged in projection along the line of sight.
{\it First Column}: Flow for ($G_h$, $r_h$) = (0.5, 5 kpc).
{\it Second Column}:  Flow for ($G_h$, $r_h$) = (2, 5 kpc).
{\it Third Column}:  Flow for ($G_h$, $r_h$) = (0.5, 20 kpc).
{\it Fourth Column}:   Flow for ($G_h$, $r_h$) = (20, 1 kpc).
Data is the same as in Figure 1.
\label{fig7}}

\vskip.1in
\figcaption[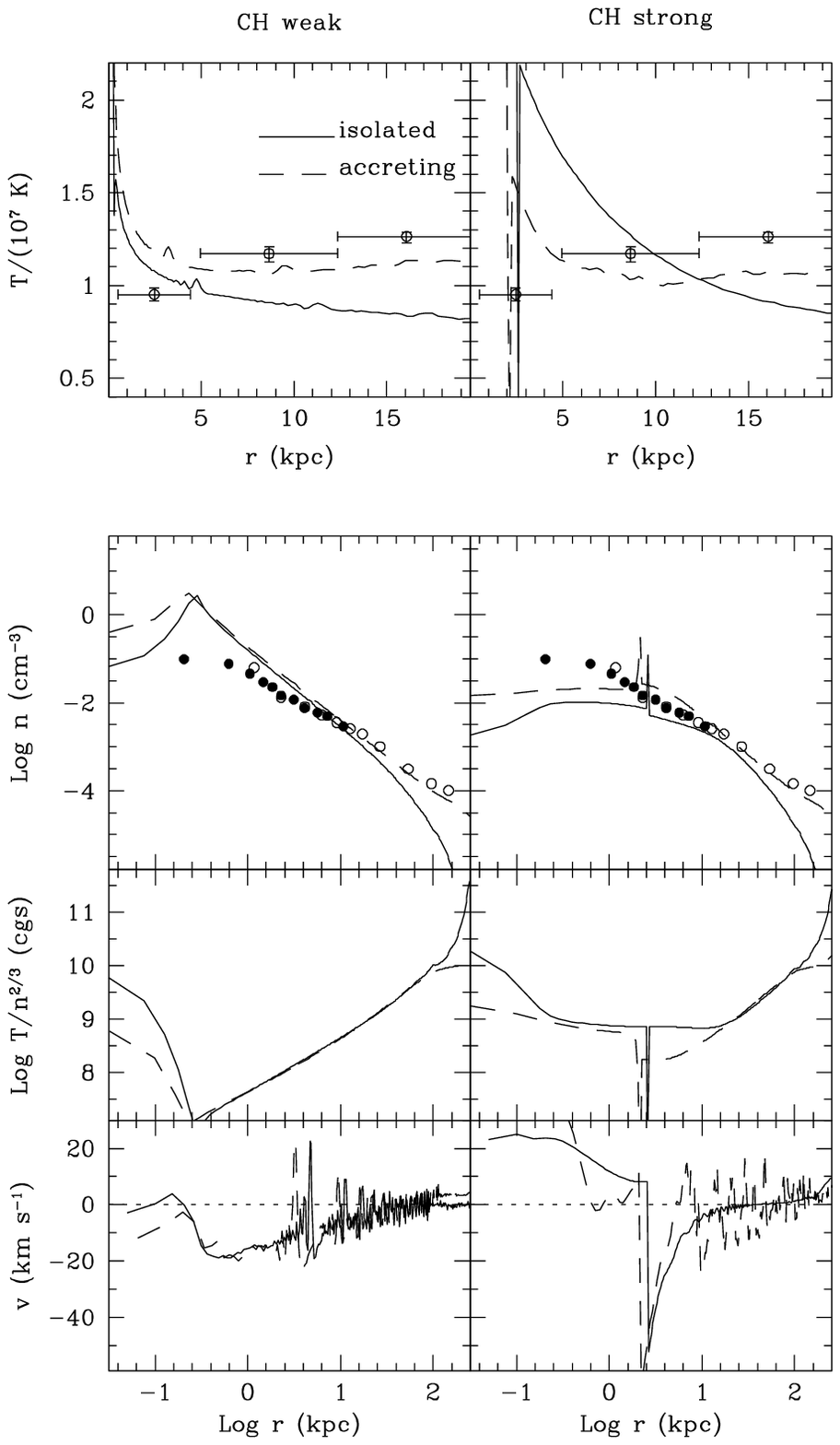]{
Gas temperature, density, specific entropy and
velocity at
time $t_n = 13$ Gyrs for two 
cooling flow models with Compton-like 
heating described by Equation (9).
Flows for isolated and accreting 
galaxies are shown with {\it solid lines} and
{\it dashed lines} respectively. 
Gas temperatures are shown as functions of physical radius, not
averaged in projection along the line of sight.
{\it Left Column}: Weakly heated flow 
with $C_h T_C = 1.6 \times 10^{43}$ 
erg cm$^2$ g$^{-1}$ s$^{-1}$.
{\it Right Column}:  Strongly heated flow
with $C_h T_C = 1.6 \times 10^{44}$.
Data is the same as in Figure 1.
\label{fig8}}

\vskip.1in
\figcaption[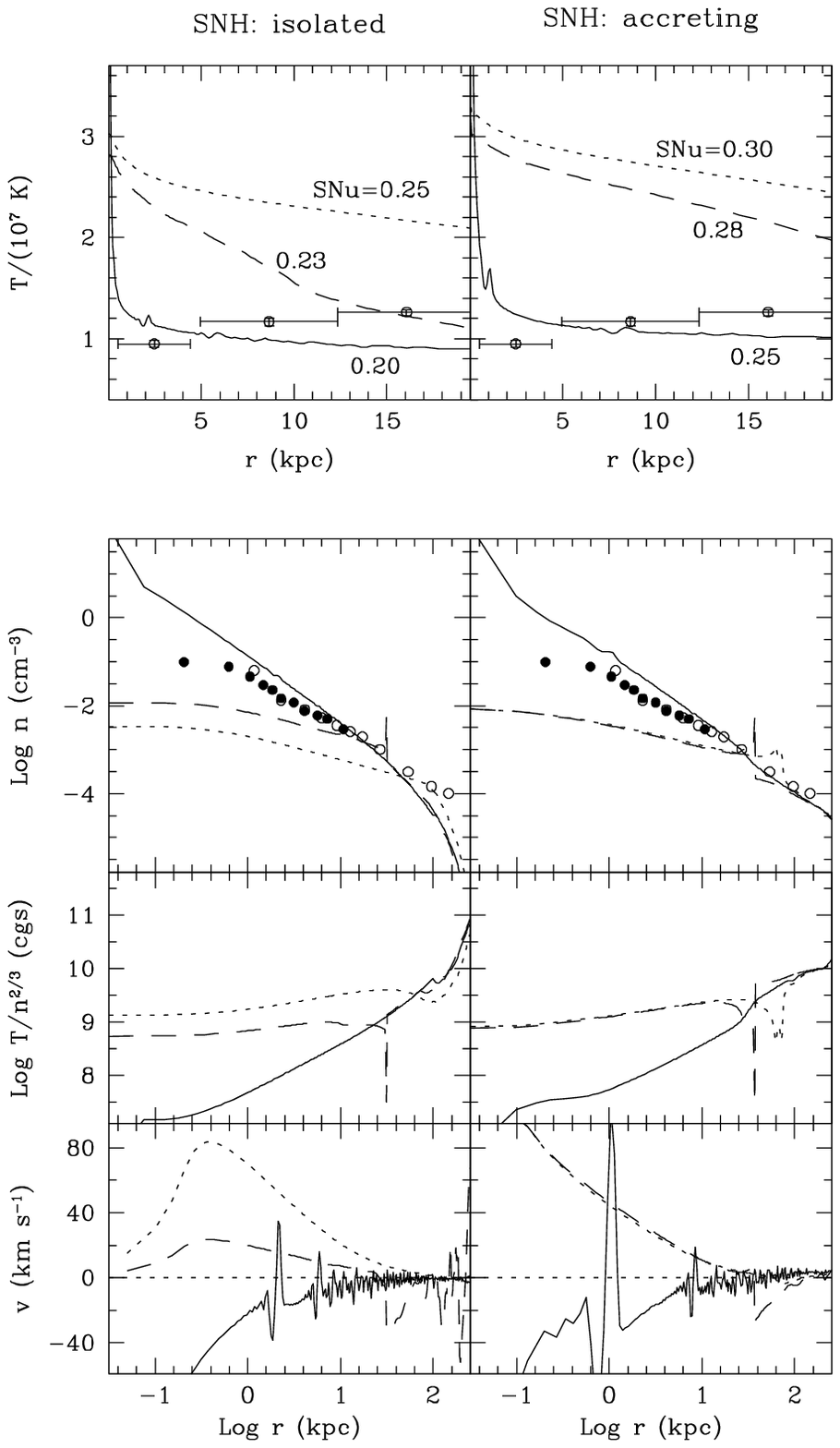]{
Representative flows that are continuously 
heated by hypothetical stellar sources, all shown at 
time $t_n = 13$ Gyrs. 
This type of SNH heating is implemented by artificially 
increasing the Type Ia supernova rate SNu$(t_n)$, 
each supernova releasing $10^{51}$ ergs. 
Values of SNu$(t_n)$ in supernovae during 100 years
per $10^{10} L_B$ label each curve in the top panel 
and apply for lower panels in each column. 
Unlike previous figures, isolated and accretion flows 
are shown in separate columns. 
Data is the same as in Figure 1.
\label{fig9}}

\vskip.1in
\figcaption[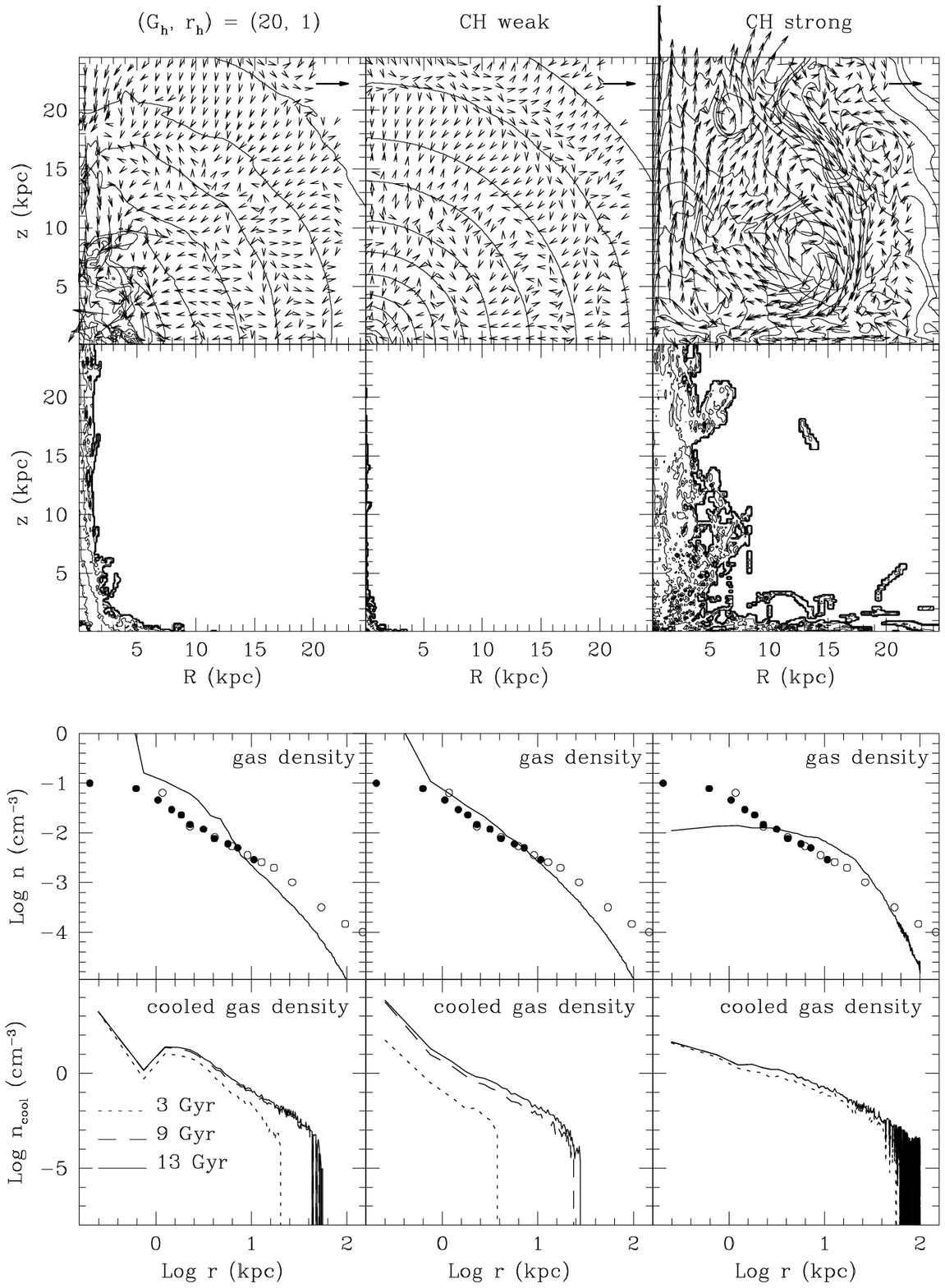]{
Three 2D flows at time $t_n = 13$ Gyrs for an isolated 
galaxy.
The uppermost panels show the gas density and velocity;
the velocity arrow at the upper right represents
200 km s$^{-1}$.
The second row of panels shows density contours 
of cooled gas. 
The third and bottom rows of panels show the 
azimuthally averaged hot gas and cooled gas densities 
respectively.
{\it Left Column}: Gaussian heated flow with
($G_h$, $r_h$) = (20, 1 kpc).
{\it Central Column}: Flow weakly heated 
with Equation (9) 
using $C_h T_C = 1.6 \times 10^{43}$ 
erg cm$^2$ g$^{-1}$ s$^{-1}$.
{\it Right Column}:  Strongly heated flow
with $C_h T_C = 1.6 \times 10^{44}$.
Data is the same as in Figure 1.
\label{fig10}}

\vskip.1in
\figcaption[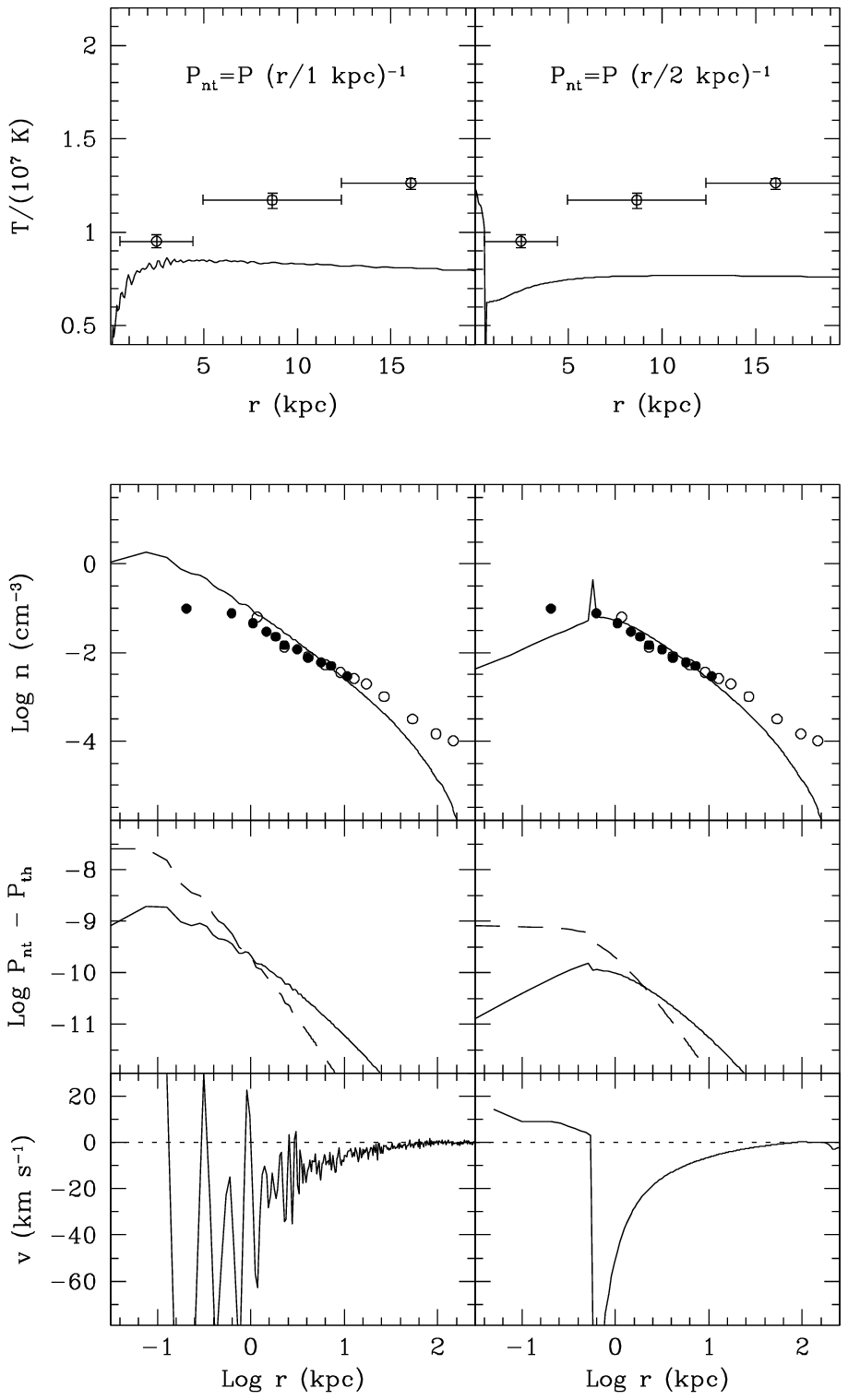]{
Gas temperature, density, pressure and 
velocity at $t_n = 13$ Gyrs  
for two unheated isolated 1D flows with additional 
non-thermal pressure described by Equation (11).
{\it Left Column}: Flow for $r_{nt} = 1$ kpc.
{\it Right Column}:  Flow for $r_{nt} = 2$ kpc.
The third pair of panels from the top compare the profiles 
of thermal pressure ({\it solid lines}) and non-thermal 
pressure ({\it dashed lines}). 
\label{fig11}}

\clearpage
\appendix

\section{Appendix}

We describe how the 1D Gaussian heated flows
with $G_h = 1$ erg gm$^{-1}$ s$^{-1}$ and
$r_h = 5$ kpc are influenced by thermal conduction
and convection.

\subsection{Influence of Thermal Conduction}

The conductive heat flux is
$$ F_{cond} = {\partial (\kappa T) \over \partial r}$$
where
$\kappa = f 1.84 \times 10^{-5} T^{5/2}/\ln{\Lambda}$
is the thermal conductivity, which can be
reduced by a factor $f \le 1$ to approximate
the inhibiting effect of tangled magnetic fields.
When the conduction becomes saturated, the flux is
replaced with
$$F_{cond,sat} = 0.4 f n k T (2 k T/ \pi m_e)^{1/2}.$$
Because $\kappa$ is very small at low temperatures,
thermal conduction can usually be ignored for cooling flows
in galaxies and galaxy groups where $T \lta 1$ keV.
However, in the flow near and inside drips
the temperature gradients are very large, increasing
the likelihood that $F_{cond}$ may alter the solution and
possibly
the mass dropout stimulated by heating at the galactic center.

The left column of
Figure A1 shows the flow variables
at $t_n = 13$ Gyr for isolated and accreting flows
with the conductivity set at the full Spitzer value, $f = 1$.
The drip that formerly appeared at 7 kpc
(right column of Figure 2) has
(temporarily) disappeared at time $t_n$
when there are no thermally unstable compression regions.
While the flows with conduction are smoother overall,
there is again an undesirable negative temperature
gradient in the central region.

A closer look at these thermally conducting
gas flows reveals
that many drips occurred at $t < t_n$ and
a large amount of gas
cooled (dropped out) at large radii.
A large mass of gas, $\sim 2.0 \times 10^{10}$
$M_{\odot}$, also cools in $r<100$ pc,
where we assume $q = 1$ to absorb the cooled gas.
Cooling dropout that occurred before
$t = 3$ Gyrs, when
$M_{cool} \sim 10^{10}$ $M_{\odot}$,
may have formed into a second generation of old stars.
By time $t_n$, $M_{cool} = 3.7 \times 10^{10}$
($7.9 \times 10^{10}$) $M_{\odot}$
has cooled
in the isolated (accretion) flow.
Most of the cooled gas
is concentrated near $r \approx 7$ kpc where the
drip appeared without conduction.

At late times, episodic drips occur in these
flows separated by $\sim 2$ Gyr periods of
relative quiescence,
resulting in a quasi-cyclic variation in the
global gas density as discussed in \S 6.2.1.
Just following a period of intense spatially
distributed mass dropout in drip waves, the hot
gas density is lower and the drip cooling
shuts down temporarily.
During the quiescent period that follows, the gas density
slowly increases due to stellar mass loss until it
becomes high enough to stimulate another episode of drip
cooling.
The high gas density in
$r \lta 10$ kpc visible in the accretion flow with
conduction occurs at a moment just prior to a period
of more intense cooling.
Two flows that appear
to be quite different at time $t_n$, such as the
isolated and accretion flows with conduction
in Figure A1
differ largely because they are momentarily
at different phases of the cooling cycle.

Since neither of the flows with thermal conduction
shown in
Figure A1 is currently undergoing drip
cooling, they would show no spectral
evidence for multiphase gas at the present time.
Nevertheless, at any particular time
about half of all E galaxies with this type of
flow should be actively
cooling and exhibit multiphase spectra.
In addition, the negative temperature gradients
would also disqualify
both solutions with full Spitzer ($f = 1$) conduction.
When these calculations were repeated with
a reduced conductivity, $f = 0.1$,
which may be more astrophysically plausible,
the drip structure and cooling dropout become almost
identical to that of the $f = 0$ case
shown in the right column of panels in Figure 2.
We conclude that thermal conduction is unable
to significantly reduce the distributed cooling
and total mass of cooled gas
in flows heated by the Gaussian $H(r,t)$.

\subsection{Influence of Convection}

Since the entropy gradient is negative
near drip waves, convection is expected.
In 1D flow the convective flux is
$$F_{conv} = 2^{-5/2} c_P g^{1/2} \rho T \ell^2 H_P^{-3/2}
\left[
\left( { d \log T \over d \log P} \right) - {\gamma - 1\over \gamma}
\right]^{3/2}$$
where $c_P = \gamma/ k /[(\gamma - 1) \mu m_p]$,
$g$ is the local acceleration of gravity,
and $H_P = P/\rho g$ is the pressure scale height
(Kippenhahn \& Weigert 1990).
This stellar expression for $F_{conv}$ is appropriate
since cooling flows are very nearly in hydrostatic
equilibrium.
We have explored various values of the mixing length
$\ell$ with similar results.
In the right column of Figure A1 we illustrate the same
Gaussian heated flow at time $t_n$ as in the right column
of Figure 2 but with convective heat flux
(and $F_{cond} = 0$) using $\ell = \max(2r/3, 3H_P/4)$,
chosen to maximize the effect of convection.
The results are very similar to the conductive case.
Frequent episodic drips still occur, producing a maximum
in the density of cooled gas at $r \sim 5$ kpc at $t_n$.
The total mass of cooled gas does not change appreciably
when convection is included.

In the right columns of
Figures 2 and A1 we also show the entropy profiles
$S(r) = (c_P/\gamma)\ln(T/ n^{2/3})$ at time $t_n$ for
Gaussian heated flows with and without convection
($G_h = 1$ erg gm$^{-1}$ s$^{-1}$ and
$r_h = 5$ kpc).
As expected, the effect of convection is to
flatten the steep entropy profile in the drip structure.
However, for the mixing length parameters we consider
the entropy gradient is still mildly negative
(in $4 \lta r \lta 20$ kpc)
even in the presence of convection.

The convective flows shown in Figures A1 --
with its unrealistically peaked temperature profile and
expected multiphase spectrum -- underscore
our conclusion: spherically symmetric
heated cooling flows cannot be brought into agreement
with XMM observations by including convection.

\vskip.1in
\figcaption[aasheatfigA1.ps]{
Gas temperature, density, specific entropy and
velocity at
time $t_n = 13$ Gyrs for 1D Gaussian heated
galactic cooling flows 
with $G_h = 1$ erg g$^{-1}$ s $^{-1}$ and $r_h = 5$ kpc.
Each panel shows flows for isolated ({\it solid lines}) and
accreting ({\it dashed lines}) galaxies.
The temperatures are shown as functions of physical radius, not
averaged in projection along the line of sight.
{\it Left Column}: Gaussian heated 
flows identical to those in the
second column of Figure 2 but including
thermal conduction at the full Spitzer value ($f = 1$).
{\it Right Column}:  Flows identical to those in the
second column of Figure 2 but including convection.
Data is the same as in Figure 1.
\label{figA1}}

\end{document}